\documentclass[twocolumn,showpacs,pre,amsmath,amssymb,superscriptaddress]{revtex4}

\usepackage{graphicx}
\usepackage{dcolumn}
\usepackage{bm}
\usepackage{amsmath}
\usepackage{amssymb}
\usepackage{mathrsfs}
\usepackage{color}

\def \rmd{\mathrm{d}}
\def \rme{\mathrm{e}}

\def \vol{\mathrm{vol}}

\def \TT{\mathbb{T}}

\newcommand\T{\rule{0pt}{0.5cm}}       
\newcommand\B{\rule[-0.3cm]{0pt}{0pt}} 

\makeindex             

\begin{document}
\title{Epi-two-dimensional fluid flow: a new topological paradigm for dimensionality}

\author{Z. Yoshida}
\address{Department of Advanced Energy, The University of Tokyo, Kashiwa, Chiba 277-8561, Japan}
\email{yoshida@ppl.k.u-tokyo.ac.jp}
\author{P. J. Morrison}
\address{Department of Physics, The University of Texas at Austin, Austin,Texas 78712, USA}
\email{morrison@physics.utexas.edu}

\date{\today}

\begin{abstract}
While a  variety of fundamental differences are known to separate two-dimensional (2D) and three-dimensional (3D) fluid  flows, it is not well understood how they are related. Conventionally,  dimensional reduction is justified by an  \emph{a priori} geometrical framework;  i.e., 2D flows occur under some geometrical constraint such as  shallowness.  However, deeper inquiry into 3D flow often finds the presence of local 2D-like structures without such a constraint, where   2D-like behavior  may be identified by the integrability of vortex lines or vanishing local helicity. Here we propose a new paradigm of flow structure  by introducing an intermediate class, termed epi-2-dimensional flow, and thereby build a topological bridge between 2D and 3D flows. 
The epi-2D property is local, and is preserved in fluid elements obeying ideal (inviscid and barotropic) mechanics;
a local epi-2D flow may be regarded as a `particle' carrying  a generalized enstrophy as its charge.
A finite viscosity may cause `fusion' of two epi-2D particles, generating helicity from their charges giving rise to 3D flow.
\end{abstract}
\maketitle


Phenomenologically, 2-dimensional (2D) fluid flow is very different from 3-dimensional (3D) flow in that the former is less-turbulent and more capable of generating  and sustaining  large-scale vortical structures\,\cite{Hasegawa1985}. This is because the dynamics of vortices in 2D systems is constrained, resulting in the suppression of  some essential mechanisms of  turbulence. Here we generalize by replacing the usual \emph{geometrical} constraint by a  \emph{topological} constraint that  extracts the essential property of 2D flow.  We call such constrained flow \emph{epi-2D}.

Conventionally, the 2D geometrical constraint is believed appropriate for a fluid with limited depth and slow
variation of physical quantities in the vertical direction compared to those in the horizontal directions.  
However, 2D-like behavior,  our {epi-2D} flow,  may occur in flows without being strictly 2D.  For example, 
sometimes in 3D systems, flow may not be totally 3D by having subdomains in which the flow is 2D-like\,\cite{CF1993}.

In this work we precisely formulate the concept of  epi-2D flow.
The difference in the  invariants of  3D and 2D systems serves as a guide:  as is well-known, the \emph{helicity} is a constant of motion in an ideal 3D flow\,\cite{Moreau,helicity}, while in 
2D geometry  the helicity degenerates to zero, being compensated by the \emph{enstrophy} (or its generalization as described below).  Usually, these two different invariants are regarded as  attributes of different dimensionality\,\cite{Fukumoto}; but,  we  switch the viewpoints and use the invariants as discriminants of  dimensionality.  Interestingly, the class of flows that conserve (appropriately generalized) enstrophy is much larger than the geometrical 2D class.  This extended class is what we refer  to as  epi-2D.   We will see that the  `fusion' of two epi-2D flows may yield a 3D flow, transmuting the corresponding enstrophies into  helicity.  In fact, the epi-2D behavior is a local property, so we can formulate a `particle picture' of transmutation.


Consider first the basic equations  and  conservation laws of fluid mechanics.
Let ${M}$ be a 3D domain containing an ideal (invicid) barotropic fluid. 
We assume ${M}=\TT^3$, the 3-torus, and ignore  boundary effects\,\cite{BC}.
Denoting by ${\rho}$ the mass density, $\bm{V}$ the fluid velocity,
$P$ the pressure,  the governing equations are 
\begin{eqnarray}
& & \partial_t{\rho} = -\nabla\cdot(\bm{V} {\rho}),
\label{mass_conservation}
\\
& & \partial_t \bm{V}=-(\bm{V}\cdot\nabla)\bm{V} - \nabla h, 
\label{momentum}
\end{eqnarray}
where  $ \rho^{-1}\nabla P = \nabla h$ with an enthalpy $h=h(\rho)$,
The energy of the system is
\begin{equation}
H = \int_{M} \left[ \frac{1}{2} |\bm{V}|^2 + \varepsilon(\rho)  \right]\,\rho\,\rmd^3 x,
\label{energy}
\end{equation}
where $\varepsilon(\rho)$ is the internal energy per unit mass  and $\partial(\rho\epsilon)/\partial\rho = h$.  
It follows by direct calculation  that the energy $H$,  the 
helicity, 
\begin{equation}
C = \int_{M} \bm{V}\cdot\bm{\omega}\,\rmd^3 x,
\label{Helicity_naive}
\end{equation}
with $\bm{\omega}=\nabla\times\bm{V}$ being  the vorticity,  
and  the total mass, $N=\int_{M} \rho\,\rmd^3x$,  are conserved.

The 2D geometrical reduction can be obtained as follows.  
Let $z$ be a `perpendicular'  coordinate in the Cartesian $(x, y, z)$ system 
and let $\bm{e}_z=\nabla z$.
The reduction with $\bm{e}_z\cdot\bm{V}=0$ and $\partial_z=0$
yields the  2D system on the $x$-$y$ plane (a flat torus $\TT^2$).
Using  $\bm{V}=( v_x, v_y, 0)^{\mathrm{T}}$ and  $\bm{v} = ( v_x, v_y)^{\mathrm{T}}$, the 
 vorticity becomes $\bm{\omega}=\nabla\times\bm{V}=\omega \bm{e}_z$, 
where $\omega=\partial_x v_y - \partial_y v_x$.
Because $\bm{V}\cdot\bm{\omega} = 0$ for this 2D reduction,  helicity conservation is now trivial:  $C\equiv0$.
Interestingly, however, a different invariant emerges:
the \emph{generalized enstrophy}
\begin{equation}
{Q} = \int_{M} f(\vartheta) \rho\,\rmd^2 x,
\label{G-enstrophy}
\end{equation}
with the \emph{potential vorticity} $\vartheta=\omega/\rho$, 
is now a constant of motion ($f$ being  an arbitrary smooth function).
It is also easy to show the constancy of the `local' enstrophy that is defined by 
replacing the domain $M$ of the integral (\ref{G-enstrophy}) 
by an arbitrary co-moving (i.e. transported by the flow $\bm{v}$) sub-domain $\Sigma(t)$.
For the  simple choice $f(\vartheta)=\vartheta$, the local enstrophy reads
${Q} = \int_{\Sigma(t)} \omega\,\rmd^2 x = \oint_{\partial\Sigma(t)} \bm{v}\cdot\,\rmd\bm{x}$, 
and the constancy of this $Q$ is known as Kelvin's circulation theorem.
For an incompressible flow ($\nabla\cdot\bm{v}=0$), we may assume $\rho=$ constant, and then
$Q$ has a special form  $\int \omega^2\,\rmd^2x $,
which is the usual {enstrophy}. 



Our formulation of  epi-2D flow  begins by inquiring into the root cause of these invariants,  deemed as a reflection of some symmetry.  In particular, we replace the geometrical symmetry 
$\bm{e}_z\cdot\, =0$ and $\partial_z=0$ that characterizes  the 2D system by a \emph{gauge symmetry} that yields an equivalent enstrophy invariant.

The fluid equations (\ref{mass_conservation})--(\ref{momentum}) are a  Hamiltonian field theory 
on a phase space $\mathcal{V}$ of fluid variables $\bm{u} = (\rho, \bm{V})^{\textrm{T}}$\,\cite{MG80,Morrison-RMP}.
The constancy of the Hamiltonian (energy) $H$ is due to $\partial_t H=0$.
In contrast, the conservation of the total mass $N$ and the helicity $C$ is independent of the choice of Hamiltonian, implying that they are not related to any explicit symmetry of the system.
Such constants of motion are called Casimir invariants. 
A possible mechanism that yields a Casimir invariant is a {gauge symmetry}
in some representation of $\mathcal{V}$. If there is  an underlying phase space $X$ of fundamental variables $\bm{\xi}$,  with the physical variables $\bm{u}$ represented by some specific combinations of $\bm{\xi}$ where the  map $\bm{\xi}\mapsto\bm{u}$ is redundant, then 
a gauge freedom occurs and  a Casimir invariant is  a Noether charge of the gauge symmetry.  
One example of this is the relabeling symmetry of the Lagrangian to Eulerian fluid representations\,\cite{pjmP96}, but for the purposes here the relevant symmetry  is that of the  \emph{Clebsch parameterization}\,\cite{Clebsch,pjmAIP,Marsden1983,Yoshida2009,YM2016}, 
for which it was shown that $N$ and $C$ are Noether charges\,\cite{TanehashiYoshida2015}.

Let $X$ be the {phase space} of Clebsch parameters
\begin{equation}
\bm{\xi}  =  (\varrho, \varphi, p, q, r, s )^{\mathrm{T}} ~\in X,
\label{canonical_variables}
\end{equation}
where each $\xi_j$ ($j=1,\cdots, 6$) is a function on the base space ${M}=\TT^3$\,\cite{suppl1}.  
On the space of \emph{observables} (i.e. smooth functionals on $X$), we define a canonical Poisson bracket
\begin{equation}
\{ F, G \} = \langle \partial_{\bm{\xi}} F, J \partial_{\bm{\xi}} G \rangle,
\label{canonical_Poisson_bracket}
\end{equation}
where $\langle \bm{\eta}, \bm{\xi} \rangle = \int_{M} \bm{\eta}\cdot\bm{\xi}\,\rmd^3x $,
$\partial_{\bm{\xi}} F$ is the gradient of $F$ in $X$,
and $J$ is the symplectic operator 
\begin{equation}
J = J_c\oplus J_c\oplus J_c,
\quad 
J_c = \left( \begin{array}{cc}
0 & I \\
-I & 0
\end{array}\right).
\label{symplectic}
\end{equation}
Given a Hamiltonian $H$,
the adjoint representation of  Hamiltonian dynamics is
${\rmd}F/{\rmd t}  = \{ F, H \}$,
which is equivalent to Hamilton's equation of motion
\begin{equation}
\partial_t \bm{\xi}= J \partial_{\bm{\xi}} H .
\label{Hamilton_eq}
\end{equation}

We relate the physical quantity $\bm{u}\in \mathcal{V}$ and $\bm{\xi}\in X$ by
$\rho\Leftrightarrow \varrho$
and (denoting $\check{p}={p}/{\varrho}$, and $\check{{r}}={{r}}/{\varrho}$)
\begin{equation}
\bm{V} \Leftrightarrow    \bm{\wp} = \nabla\varphi + \check{p} \nabla q + \check{{r}} \nabla {s} .
\label{Clebsch}
\end{equation}
Writing a vector as (\ref{Clebsch}) is called the  \emph{Clebsch parameterization}\,\cite{Clebsch,pjmAIP,Marsden1983}.
The five Clebsch parameters $(\varphi,  \check{p}, q,  \check{r}, s)$ are sufficient to represent every 3-vector (1-form in 3D space)\,\cite{Yoshida2009}.
Inserting (\ref{Clebsch}) into the fluid energy (\ref{energy}), we obtain the  Hamiltonian
\begin{equation}
H(\bm{\xi}) = \int_{M} \left[ \frac{1}{2} {\left| \nabla\varphi + \frac{p}{\varrho} \nabla q 
+\frac{{r}}{\varrho} \nabla {s}\right|^2} + \varepsilon(\varrho)  \right] \varrho \,\rmd^3 x.
\label{Hamiltonian}
\end{equation}
With this $H$, the equation of motion (\ref{Hamilton_eq}) reads\,\cite{BC}
\begin{equation}
\left\{
\begin{array}{l}
\partial_t \varrho + \nabla\cdot(\bm{V} \varrho) =0, \\
\partial_t \varphi + \bm{V}\cdot\nabla\varphi = h- \frac{1}{2}{V^2}, \\
\partial_t p + \nabla\cdot(\bm{V} p) =0, \quad
\partial_t q + \bm{V}\cdot\nabla q =0, \\
\partial_t r + \nabla\cdot(\bm{V} r) =0 , \quad
\partial_t s + \bm{V}\cdot\nabla s =0 .
\end{array} \right.
\label{H-1} 
\end{equation}
The first equation of (\ref{H-1}) is nothing but the mass conservation law (\ref{mass_conservation}).
Evaluating $\partial_t\bm{V}$ by inserting (\ref{Clebsch}) and using (\ref{H-1}), we
obtain (\ref{momentum}).
Hence, Hamilton's equation (\ref{Hamilton_eq}) with the Hamiltonian (\ref{Hamiltonian})
describes  fluid motion obeying (\ref{mass_conservation}) and (\ref{momentum})\,\cite{Lin,Zakharov,pjmAIP}.
The invariance of the helicity $C$ 
also follows from (\ref{H-1}).

We examine how the 2D geometrical reduction works in the Hamiltonian formalism.
In a 2D system (${M} = \TT^2$), we can parameterize a general 2D velocity as
\begin{equation}
\bm{V} \Leftrightarrow  \bm{\wp} = \nabla \varphi + \check{p}  \nabla q ,
\quad (\check{p} =p/\varrho).
\label{Clebsch-2D}
\end{equation}
Here, only three Clebsch parameters $\varphi$, $\check{p}$, and $q$ are sufficient\,\cite{Yoshida2009}.
The vorticity is $\omega=  (\nabla \check{p} \times  \nabla q)\cdot\bm{e}_z$.
The helicity cannot be defined  in the 2D space.
The potential vorticity is a scalar $\vartheta=\omega/\varrho$,
and the generalized enstrophy reads $Q=\int_{\Sigma(t)} f(\omega/\varrho) \varrho\,\rmd^2x$,
where a sub-domain $\Sigma(t)$ is moved by the group-action of $\mathrm{e}^{t\bm{v}}$.
We can easily verify $\rmd Q/\rmd t=0$ by  (\ref{H-1}).



Epi-2D flow is obtained in the 3D setting with the phase space $X$ on the base space $M=\TT^3$ by  setting ${r}=0$\,\cite{2.5D}.  
The corresponding physical fields are $\rho\Leftrightarrow \varrho$ and 
\begin{equation}
\bm{V} \Leftrightarrow  \bm{\wp} =  \nabla \varphi + \check{p}  \nabla q ,
\quad \left(\check{p}= {p}/{\varrho}\right) .
\label{epi-2D-2}
\end{equation}
This yields a  2D-like representation, but there is a  difference between (\ref{Clebsch-2D}) for 
2D flow and (\ref{epi-2D-2}) for  epi-2D flow, for the latter resides in the 3D domain $\TT^3$.

Epi-2D flow is generated by the  reduced Hamiltonian 
\begin{equation}
H(\bm{\xi}) = \int_{M} \left[ \frac12 {\left| \nabla \varphi + \frac{p}{\varrho} \nabla q \right|^2 }  + \varepsilon(\varrho)  \right] \varrho\,\rmd^3 x, 
\label{reduces-Hamiltonian}
\end{equation}
giving  the  3D  equations  (\ref{mass_conservation}) and (\ref{momentum}).
While ${s}$ does not appear in  (\ref{reduces-Hamiltonian}),
it obeys  the same equation (\ref{H-1}) but 
with $\bm{V}$ independent of ${s}$.
Such a field, co-moving with the epi-2D flow,
is called a \emph{phantom} field\,\cite{YoshidaMorrisonFDR2014}.
Or,  $s$ is a gauge field with the observables  blind to its initial value.

As previously remarked,  (\ref{epi-2D-2})  cannot represent  an arbitrary 3D flow:  
epi-2D flow may have a finite vorticity $\bm{\omega}=\nabla\times\bm{\wp}=\nabla\check{p}\times\nabla q$, but its 
helicity density
$\bm{\wp}\cdot(\nabla\times\bm{\wp}) 
= \nabla\varphi\cdot(\nabla\check{p}\times\nabla q) 
= \nabla\cdot(\varphi \nabla\check{p}\times\nabla q )$ 
is an exact differential, implying  zero helicity,  $C=\int_{{M}} \bm{\wp}\cdot(\nabla\times\bm{\wp}) \,\rmd^3x =0$.

As for 2D, this degeneracy of the helicity is compensated by 
a different invariant 
obtained by  extending  the generalized enstrophy to  3D.
With  the phantom $s$, 
\begin{equation}
{Q} := \int_{\Omega(t)} f(\vartheta)\, \varrho\,\rmd^3 x,
\quad \vartheta = \frac{\bm{\omega} \cdot \nabla{s}}{\varrho}, 
\label{enstrophy-2}
\end{equation}
with  arbitrary $f$ and an arbitrary co-moving 3D volume element $\Omega(t)\subset{M}$,  is seen to be conserved  upon  making use of  (\ref{H-1}). 
If we  choose $s=z$,  then (\ref{enstrophy-2})  reduces to the 2D  form  $Q=\int_{\Sigma(t)} f(\omega/\varrho) \varrho\,\rmd^2 x$.
In what follows, we choose the simplest case $f(\vartheta) =\vartheta$.



\emph{Local}  epi-2D regions within a 3D flow can be exploited  to define particle-like behavior.
With  the general 3D parameterization
$\bm{V} \Leftrightarrow  \bm{\wp} = \nabla\varphi + \check{p}\nabla q + \check{{r}}\nabla {s}$,
a region in which $\check{{r}}=0$ may be called an  \emph{epi-2D domain}.
Since $\check{{r}}$ co-moves with the fluid,
every infinitesimal volume element, say  $\Omega_j(t)$ with elements  indexed  by  $j$, 
included in an \emph{epi-2D domain} may be viewed as a {quasiparticle},
which we call an \emph{epi-2D particle}.
The generalized enstrophy evaluated for the vorticity
$\bm{\omega}_+=\nabla\check{{p}}\times\nabla{q}$ in $\Omega_j(t)$, denoted by ${Q}_+(\Omega_j)$, is a constant of motion.
Here  the index `+' is used to distinguish from  counterpart domains where $\check{p}=0$, for which 
${Q}_-(\Omega_j) = \int_{\Omega_j(t)} \vartheta_- \varrho\,\rmd^3x$,  where 
$\vartheta_- = (\bm{\omega}_-\cdot\nabla q)/\varrho$ with  
the vorticity $\bm{\omega}_- = \nabla\check{{r}}\times\nabla{s}$.

We call ${Q}_\pm (\Omega_j)$ the \emph{charge} of the {epi-2D particle} $\Omega_j$.
It is remarkable that both ${Q}_+$ and ${Q}_-$ are invariant even in 3D flows.
Both charges  essentially measure  the `circulations' of the decomposed components of the flow (cf.\ 
the  comment following (\ref{G-enstrophy})).
In fact, Kelvin's circulation theorem applies in general 3D flows.
The merit of the use of the charges $Q_\pm$, in comparison with the conventional circulation,
is in that they can delineate the local flow structure.
As far as a particle (volume element) carries only $Q_+$ or $Q_-$, it is epi-2D.
However, when the vorticity is in a \emph{mixed state}, i.e.,
$\bm{\omega} = \bm{\omega}_++\bm{\omega}_-= \nabla \check{p}\times\nabla q + \nabla\check{{r}}\times\nabla{s}$,
the particle becomes 3D,  and then,   helicity is created from the charges $Q_+$ and $Q_-$.
Indeed, the integrands of $Q_+$ and $Q_-$ (the charge densities) are, respectively, 
\[
\mathscr{Q}_+ 
=  (\nabla\check{{p}}\times\nabla q)\cdot\nabla{s},
\quad 
\mathscr{Q}_- 
=  (\nabla\check{{r}}\times\nabla{s})\cdot\nabla q ,
\]
while the integrand of $C$ (the helicity density) is
\begin{equation}
\mathscr{C} = \bm{\wp}\cdot(\nabla\times\bm{\wp})= \check{{r}}\mathscr{Q}_+ + \check{{p}}\mathscr{Q}_- + \mathscr{C}_{\textrm{ex}} ,
\label{helicity_density}
\end{equation}
where $\mathscr{C}_{\textrm{ex}}= \nabla\varphi\cdot(\nabla\check{{p}}\times\nabla{q}+\nabla\check{{r}}\times\nabla{s})$
is the exact part of the helicity density.
The residual helicity density 
$\mathscr{C}_{\mathrm{r}}=\mathscr{C}-\mathscr{C}_{\textrm{ex}} = \check{{r}} \mathscr{Q}_+ + \check{{p}}\mathscr{Q}_-$
describes the coupling of epi-2D particles; evidently, either $\check{r}=0$ or $\check{p}=0$ causes  $\mathscr{C}_{\mathrm{r}}$ to vanish.
Conversely, the combination of two charges $\mathscr{Q}_+$ and $\mathscr{Q}_-$ yields $\mathscr{C}_{\mathrm{r}}$.
Therefore, $\mathscr{C}_{\mathrm{r}}=0$ can be used as an alternative definition of epi-2D flow.

\begin{figure}[tb]
\begin{center}
\includegraphics[scale=0.25]{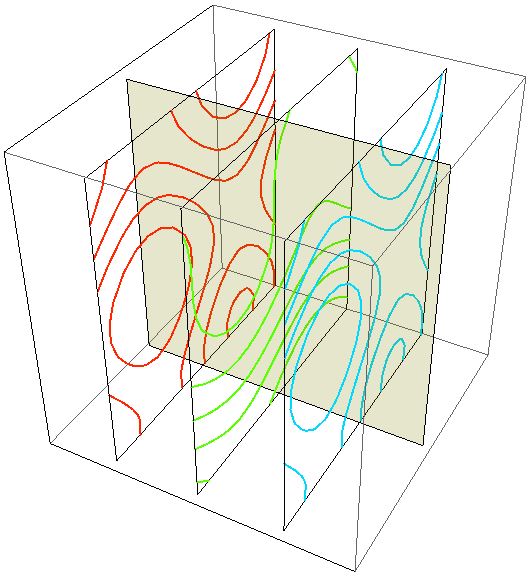}~
\includegraphics[scale=0.28]{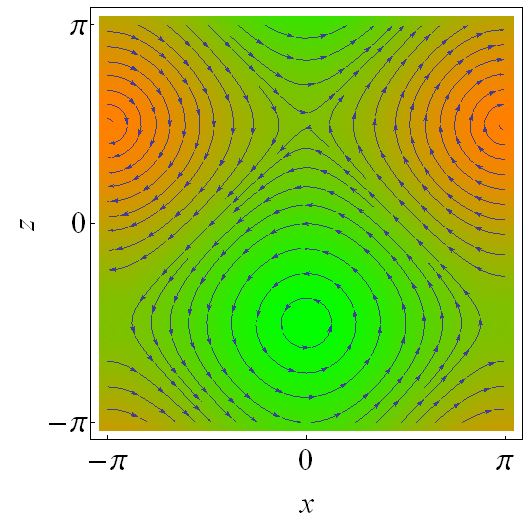}
\caption{
\label{fig:epi2D}
An epi-2D flow given by $\bm{V}_+$.  
(Left) The integrable vortex lines.
(Right) Contours of  $\mathscr{Q}_+$ and the flow vector on the surface $y=0$
indicated by the gray cross-section in the left figure
(color code ranges from orange$=2$ to green$=-2$).
}
\end{center}
\end{figure}

\begin{figure}[htb]
\begin{center}
\includegraphics[scale=0.25]{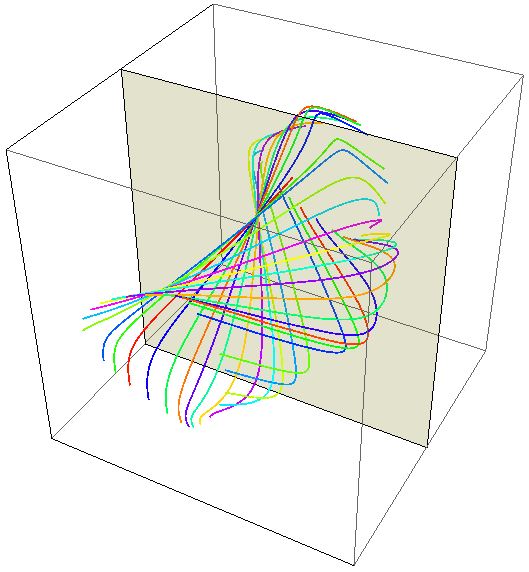}~
\includegraphics[scale=0.28]{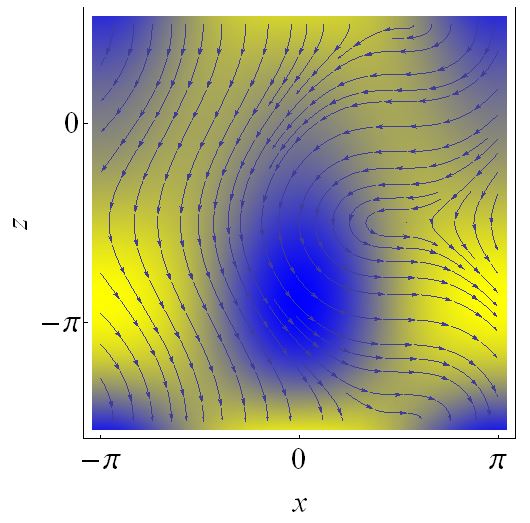}
\caption{
\label{fig:ABC}
The combination $\bm{V}=\bm{V}_++\bm{V}_-$ yields a 3D flow.
(Left) The vortex lines become chaotic (non-integrable).
(Right) Contours of $\mathscr{C}_{\mathrm{r}}$ 
together with the surface-aligned component of the flow vector on the surface $y=0$
indicated by the gray cross-section in the left figure
(color code ranges from blue$=5$ to yellow$=-1$).
}
\end{center}
\end{figure}

We give an illustrative example to visualize epi-2D flow and its transition to 3D.
Let $(x, y, z)$ be  Cartesian coordinates, and $\bm{V} = \alpha \bm{V}_+ + \beta \bm{V}_- $ with
\[
\bm{V}_+ = \left( \begin{array}{c}
b\sin y - c\cos z \\
0 \\
a\sin x 
\end{array} \right) ,
~
\bm{V}_- = \left( \begin{array}{c}
0 \\
c\sin z - a\cos x \\
 - b\cos y
\end{array} \right) ,
\]
where $\alpha, \beta, a, b$ and $c$ are arbitrary real constants, and the  vorticities satisfy $\bm{\omega}_\pm=\nabla\times\bm{V}_\pm=\bm{V}_\mp$. 
When $\alpha=\beta =1$, $\bm{V}$ is the famous ABC flow\,\cite{abc}, satisfying $\nabla\times\bm{V}=\bm{V}$.
We may cast $\bm{V}$ into the Clebsch form (\ref{Clebsch}) with
\begin{equation}
\left\{ \begin{array}{ll}
\varphi = (\alpha a\sin x - \beta b\cos y) z, & ~
\\
\check{p} = \alpha[b\sin y - c\cos z -(a\cos x) z], & q=x,
\\
\check{{r}} = \beta[ c\sin z-a\cos x  - (b\sin y) z], & s=y.
\end{array} \right.
\label{ABC-Clebsch}
\end{equation}
When $\alpha=1$ and $\beta=0$, we obtain an epi-2D flow $\bm{V}_+$ with 
$\mathscr{Q}_+=c\sin z - a\cos x$.
The co-presence of $\bm{V}_+$ and $\bm{V}_-$ creates a 3D flow;
when $\alpha=\beta=1$, $\bm{V}$ has the residual helicity density
$\mathscr{C}_{\mathrm{r}} = {r} \mathscr{Q}_+ + p \mathscr{Q}_-$.
In Fig.\,\ref{fig:epi2D}, we show the structure of $\bm{V}_+$ and the distribution of $\mathscr{Q}_+$ on a surface $y=0$.
Figure\,\ref{fig:ABC} depicts the 3D flow $\bm{V}=\bm{V}_++\bm{V}_-$ and $\mathscr{C}_{\mathrm{r}}$ on a cross-section of $y=0$.


\begin{table}[tb]
\caption{Topological classification of flows. 
$\bm{V}$: 3D flow, $\bm{V}_{\mathrm{r}}=\bm{V}-\nabla\varphi$: solenoidal component,
$\mathscr{Q}_\pm$: generalized enstrophy density,
$Q_\pm=\int_{\Omega(t)}\mathscr{Q}_\pm\,\rmd^3x$: generalized enstrophy,
$\mathscr{C} $: helicity density, 
$\mathscr{C}_{\mathrm{r}}$: residual helicity density, $C=\int_M \mathscr{C}_{\mathrm{r}}\,\rmd^3x$: helicity.
Notice that the local integral $\int_{\Omega(t)} \mathscr{C}_{\mathrm{r}}\,\rmd^3x$ is not a constant of motion.
}
\begin{center}
\begin{tabular}{l|l|l}
\hline
classification & representation  & invariants 
\\ \hline
$\begin{array}{l} \textrm{vorticity-free} \\ (\nabla\times\bm{V}=0) \end{array}$ & $\bm{V}=\nabla\varphi$ &
 $\begin{array}{l} \mathscr{C}=0 \\ \mathscr{Q}=0 \end{array}$ 
\\ \hline
$\begin{array}{l} \textrm{helicity-free} \\ (\bm{V}\cdot\nabla\times\bm{V}=0) \end{array}$ & $\bm{V}=\check{p}\nabla q$ &
 $\begin{array}{l} \mathscr{C}=0 \\ {Q}_\pm \end{array}$ 
\\ \hline
$\begin{array}{l} \textrm{epi-2D} \\(\bm{V}_{\mathrm{r}}\cdot\nabla\times\bm{V}_{\mathrm{r}}=0) \end{array} $ & $\bm{V}=\nabla\varphi+\check{p}\nabla q$ &
 $\begin{array}{l} \mathscr{C}_{\mathrm{r}}=0 \\ {Q}_\pm \end{array}$
\\ \hline
~general & $\bm{V}=\nabla\varphi+\check{p}\nabla q+ \check{{r}} \nabla{s}$ & 
$\begin{array}{l} C \\ {Q}_\pm \end{array}$
\\
\hline
\end{tabular}
\end{center}
\label{table:hierarchy}
\end{table}

Table\,\ref{table:hierarchy} summarizes our newly proposed classification of flows.
The smallest set hosts vorticity-free, potential flow (or \emph{lamellar} field\,\cite{lamellar}). 
The next of the hierarchy includes `weighted' potential flows (or \emph{complex lamellar} field) 
that  have vorticity but still zero helicity density.
A further generalization yields the epi-2D flows 
that have only exact helicity density (hence, $\mathscr{C}_{\mathrm{r}}=0$).
The epi-2D class subsumes conventional 2D systems
where we may take $s=z$ (the perpendicular coordinate);
this is possible since $\bm{\omega}$ is aligned to the fixed vector $\bm{e}_z$\,\cite{symmetries}.
As the generalization of the \emph{a priori} base space of a 2D system,
an epi-2D flow has intrinsic vortex surfaces (cf.\,\cite{Frobenius}).
While the direction of  $\bm{\omega}$ changes dynamically,
the vortex lines remain integrable, keeping the similarity to 2D flows  (cf.\,\cite{CF1993}).
Contrary to 2D flow, however, epi-2D flow allows for vortex stretching,
which may make the epi-2D particle thinner (thus, the possibility of singularity generation is not precluded).
A general 3D flow may be viewed as a mixed state of epi-2D particles, with 
each particle carrying  a charge of  ${Q}_+$ or ${Q}_-$\,\cite{suppl2}.
When particles with ${Q}_+$ and ${Q}_-$ occupy a same volume element, they produce a helicity to make the volume 3D  (cf.\,\cite{3D} for experimental visualization of knotted vortices).
Otherwise, the volume is epi-2D.
The epi-2D property is topologically invariant; i.e., an epi-2D volume remains so under  ideal fluid motion.
If some non-ideal process, such as vortex reconnection, occurs\,\cite{3D2,3D3,3D4},
however, two epi-2D particles can fuse to generate  helicity. 

The class of locally epi-2D flows is capable of describing strongly heterogeneous 3D vortex dynamics
where the helicity density is localized in narrow subdomain 
(such local structures often manifest as coherent vortices, and are called \emph{worms})\,\cite{worm}.
We note that vortex stretching can happen in any of these subdomains.   

In conclusion, the newly formulated epi-2D vector fields are useful for delineating between  
mixed states of order and disorder,
which indeed appear as intermittency, coherent vortices, or various local structures in fluid systems.
Previously, the framework of 2D geometry was the only one for describing  
simple (integrable) vortex structures and discussing their moderate (or ordered) dynamics.
However, such structures/dynamics can manifest themselves without this \emph{a priori} geometrical constraint;
they are more flexible and ubiquitous in general 3D space, as we do observe in actual phenomena.
The epi-2D class abstracts the \emph{topological} characteristics of the usual 2D flows;
it persists under deformations by ideal fluid motion  (including stretching); 
being a local property, it is suitable for characterizing the mixture of epi-2D and true 3D dynamics;
it bridges 2D and 3D by elucidating how 3D flow is created from the epi-2D prototype, or conversely,
how epi-2D degenerates to 2D.
Here we discussed fluid mechanics, but the paradigm of Table\,\ref{table:hierarchy} for  3D vectors
applies to a variety of fields, including magnetic fields\,\cite{magnetic}, optical vortices\,\cite{optic},
as well as chiral charge-density waves\,\cite{Ishioka2010}.

ZY was supported by JSPS KAKENHI contract 15K13532 and 17H01177,
and PJM  by USDOE  DE-FG02-04ER-54742. PJM  thanks  George Miloshevich for useful suggestions. 




\newpage

\begin{center}
{\Large \it Appendix}
\\
\bigskip
\end{center}

\appendix

\section{Boundary conditions}
Here we formulate the ideal fluid system occupying a smoothly bounded compact domain $\Omega\subset\mathbb{R}^3$,
and go into the problems pertaining to the boundary.
The boundary condition we impose is
\begin{equation}
\bm{n}\cdot\bm{V}=0 \quad \mathrm{on}~\partial\Omega
\label{BC}
\end{equation}
 ($\bm{n}$ the unit normal vector onto the boundary $\partial\Omega$).
This is a physically acceptable assumption implying the `confinement' of the fluid inside the boundary.
While (\ref{BC}) poses a restriction on $\bm{V}$, it also influences $\rho$ (or the enthalpy $h(\rho)$); 
the compatibility of (\ref{BC}) and (2) demands
\begin{equation}
\bm{n}\cdot\nabla h(\rho) = -\bm{n}\cdot(\bm{V}\cdot\nabla)\bm{V} \quad \mathrm{on}~\partial\Omega .
\label{BC-2}
\end{equation}

\bigskip
\noindent
\textit{Functional derivatives}

When we evaluate $\partial_{\bm{\xi}} H$, we need to eliminate the boundary terms upon integration by parts.
For the periodic domain $M=\mathbb{T}^3$, boundary terms vanish automatically.  
But here, we need the boundary condition (\ref{BC}).

Let us demonstrate the derivation of the same canonical system (12) for the Hamiltonian (11) defined on a bounded domain $\Omega$.
The gradient $\partial_{\bm{\xi}} H(\bm{\xi})$ is defined by
\begin{equation}
H(\bm{\xi}+\epsilon\tilde{\bm{\xi}}) - H(\bm{\xi})
= \epsilon \langle \partial_{\bm{\xi}} H(\bm{\xi}), \tilde{\bm{\xi}} \rangle + O(\epsilon^2) 
\label{gradient}
\end{equation}
for every $\tilde{\bm{\xi}}$ satisfying the boundary condition.
The variations for which we need the boundary condition (\ref{BC}) are those of $\tilde{\varphi}$, $\tilde{q}$, and $\tilde{s}$.
For example, we observe (omitting $O(\epsilon^2)$ terms)
\begin{eqnarray*}
& & H(\varphi+\epsilon\tilde{\varphi}) - H(\varphi) 
\\
& & ~~= \epsilon \int_\Omega (\bm{V}\cdot\nabla\tilde{\varphi} )\varrho\,\rmd^3x
\\
& & ~~=
- \epsilon \int_\Omega \tilde{\varphi} \, \nabla\cdot(\bm{V} \varrho) \,\rmd^3x 
+ \epsilon \int_{\partial\Omega} \tilde{\varphi} \varrho (\bm{n}\cdot\bm{V}) \,\rmd^2 x.
\end{eqnarray*}
Using (\ref{BC}), we obtain
$\partial_{\varphi} H = - \nabla\cdot(\bm{V} \varrho) $.
By similar calculations, we can derive (12).

\bigskip
\noindent
\textit{Characteristics of the hyperbolic system}

The ideal fluid equations (1)-(2), or their Clebsch parameterized forms (12), constitute a nonlinear hyperbolic system.
For an arbitrary vector $\bm{V}$, its streamlines are determined by
\begin{equation}
\frac{\rmd \bm{x}}{\rmd t} = \bm{V}(\bm{x}) .
\label{characteristicODE}
\end{equation}
If $\bm{V}$ satisfies (\ref{BC}), the boundary $\partial\Omega$ parallels the streamlines.
By (12), the Clebsch parameters $\check{p}, q, \check{r}$ and $s$ are constant along the streamlines
(i.e., they are Lie-dragged scalars).
Once an initial condition is given, their boundary values are automatically determined by the evolution equation (12).
The boundary values of the remaining two parameters, $\varrho$ and $\varphi$ must be determined  by 
the boundary conditions (\ref{BC}) and (\ref{BC-2}), which read
\begin{eqnarray}
\bm{n}\cdot\nabla\varphi &=& -\bm{n}\cdot( \check{p}\nabla q + \check{r}\nabla s) \quad \mathrm{on}~\partial\Omega,
\label{BC-Clebsch}
\\
\bm{n}\cdot\nabla h(\varrho) &=& -\bm{n}\cdot(\bm{V}\cdot\nabla)\bm{V} \quad ~~~~\mathrm{on}~\partial\Omega .
\label{BC-Clebsch-2}
\end{eqnarray}
These two equations account for the Neumann boundary conditions for $\varrho$ and $\varphi$.
The characteristics for $\varrho$ and $\varphi$ are those of sound waves 
(propagating on the fluid moving with the velocity $\bm{V}$);
hence $\partial\Omega$ is non-characteristic.
Physically, the interaction of the sound waves with the boundary modifies the the enthalpy $h(\varrho)$ and the velocity $\bm{V}$
so that (\ref{BC-Clebsch})-(\ref{BC-Clebsch-2}) are satisfied.
The simultaneous equations (\ref{BC-Clebsch}) and (\ref{BC-Clebsch-2}),
together with the characteristic equation (\ref{characteristicODE}) with unknown $\bm{V}$ constitute 
a rather involved system.
In comparison with their na\"ive form (\ref{BC})-(\ref{BC-2}), however,
the two components that are controlled by the boundary conditions are specified by the Clebsch parameterization.
The following incompressible model has a more transparent structure.
 
\bigskip
\noindent
\textit{Incompressible model}

In the incompressible model, the sound waves are removed, and thus only the streamlines of the flow $\bm{V}$ remain
as the characteristics.
By (\ref{BC}), the boundary $\partial\Omega$ parallels the characteristic curve,
i.e., $\partial\Omega$ is characteristic.
Here we show how the autonomous evolution of the Clebsch parameters maintain the compatibility with the
boundary condition (\ref{BC-Clebsch}).

In the incompressible system, $\varrho$ is not dynamical (we assume $\varrho=1$).
Accordingly, the conjugate variable $\varphi$ is separated from the phase space;
we consider a reduced phase space:
\begin{equation}
\bm{\xi}_{ic} = (p,q,r,s)^{\textrm{T}} \in X_{ic}.
\label{IC-phase_space}
\end{equation}
Notice that $\check{p}=p$ and $\check{r}=r$, because $\varrho=1$.
The Hamiltonian is written as a function of the vorticity $\bm{\omega}=\nabla p \times \nabla q + \nabla r \times \nabla s$:
\begin{equation} 
H(\bm{\xi}_{ic}) = \int_\Omega \frac{1}{2}\left| \mathrm{curl}^{-1} (\nabla p \times \nabla q + \nabla r \times \nabla s) \right|^2\,\rmd^3x.
\label{Hamiltonian-ic}
\end{equation}
Here we define the de-curling operator $\mathrm{curl}^{-1}$ as
\begin{equation}
 \mathrm{curl}^{-1} (\nabla p \times \nabla q + \nabla r \times \nabla s) = \nabla\phi + p\nabla q + r\nabla s
 \label{de-curl-1}
\end{equation}
with $\phi$ determined (for each time) by solving an elliptic PDE
\begin{equation}
\left\{ \begin{array}{ll}
\Delta \phi = - \nabla\cdot (p\nabla q + r\nabla s) & \mathrm{in}~\Omega ,
\\
\bm{n}\cdot\nabla\phi = - \bm{n}\cdot (p\nabla q + r\nabla s) & \mathrm{on}~\partial\Omega .
\end{array} \right.
\label{de-curl-2}
\end{equation}
Evidently, $\bm{V}=\mathrm{curl}^{-1} (\nabla p \times \nabla q + \nabla r \times \nabla s) $
satisfies the incompressibility $\nabla\cdot\bm{V}=0$ and the boundary condition (\ref{BC}).

\begin{table*}[htb]
\label{table:translation}
\caption{The correspondence between the notations of differential geometry and those of vector analysis.  
Here the base space is a 3D manifold $M$.
We denote by $\{ \rmd x^1, \rmd x^2, \rmd x^3 \}$ the basis of the cotangent bundle $T^* M$,
and identify $\rmd x^j$ with a basis vector $\bm{e}^j$ ($j=1,2,3$).
We denote the tangent vector by $V = V^1 \partial_{x^1} + V^2 \partial_{x^2} +V^3 \partial_{x^3} \in T M$.
}
\begin{center}
\begin{tabular}{l|l|l}
\hline
differential forms & form notation & vector notation 
\\ \hline
0-form \T\B & 
$f$ & $f$
\\
1-form \T\B & $u = u_1 \rmd x^1 + u_2 \rmd x^2 + u_3 \rmd x^3$ & 
$\bm{u}= u_1\bm{e}^1 + u_2\bm{e}^2+u_3\bm{e}^3$
\\
2-form \T\B & $\omega = \omega_1 \rmd x^2\wedge\rmd x^3 + \omega_2 \rmd x^3\wedge\rmd x^1+\omega_3 \rmd x^1\wedge\rmd x^2  $ & 
$\bm{\omega}= \omega_1\bm{e}^1 + \omega_2\bm{e}^2+\omega_3\bm{e}^3$
\\
3-form \T\B & $\varrho=\rho \rmd x^1\wedge\rmd x^2\wedge\rmd x^3$ & $\rho$
\\ \hline \hline
exterior derivatives & ~ & ~
\\ \hline
0-form $\rightarrow$ 1-form \T\B & 
$\rmd f = \frac{\partial f}{\partial x^j}~ \rmd x^j$ & $\nabla f$
\\
1-form $\rightarrow$ 2-form \T\B & $\rmd u = \frac{\partial u_j}{\partial x^k} ~\rmd x^k\wedge\rmd x^j $ & $\nabla\times\bm{u}$ 
\\
2-form $\rightarrow$ 3-form \T\B & $\rmd \omega =  \frac{\partial \omega_j}{\partial x^j} ~ \rmd x^1\wedge\rmd x^2\wedge\rmd x^3 $ & $\nabla\cdot\bm{\omega}$ 
\\
\hline \hline
interior products & ~ & ~
\\ \hline
1-form $\rightarrow$ 0-form \T\B & $i_V u = V^j u_j$ & $\bm{V}\cdot\bm{u}$
\\
2-form $\rightarrow$ 1-form \T\B & $i_V \omega=- \epsilon_{ijk} V^i \omega_j ~\rmd x^k $ & $-\bm{V}\times\bm{\omega}$ 
\\
3-form $\rightarrow$ 2-form \T\B & $i_V \varrho = \epsilon_{ijk} V^i \rho ~\rmd x^j \wedge \rmd x^k $ & $\bm{V} \rho$
\\
\hline
\end{tabular}
\end{center}
\label{table:hierarchy}
\end{table*}

The reduced system of canonical equations are
\begin{equation}
\left\{
\begin{array}{l}
\partial_t p +  \bm{V}\cdot\nabla p =0, \quad
\partial_t q + \bm{V}\cdot\nabla q =0, \\
\partial_t r +  \bm{V}\cdot\nabla r =0 , \quad
\partial_t s + \bm{V}\cdot\nabla s =0 ,
\end{array} \right.
\label{H-ic} 
\end{equation}
for which the boundary $\partial\Omega$ is characteristic; hence we do not control $\bm{\xi}_{ic}$ at the boundary.
Notice that the characteristics of the sound waves have been removed;
instead, the elliptic equation (\ref{de-curl-2}) determines $\phi$ to satisfy the boundary condition (\ref{BC}).

\vspace*{1cm}
 
\section{Differential-geometrical formulation}
Here we put the formulation of fluid mechanics into the perspective of differential geometry. 
By doing so, we can elucidate the deeper structure underlying the Clebsch parameterization.

We start by refining the definition of the phase space $X$; see (6).
The Clebsch parameters are
\begin{equation}
\bm{\xi}  =  (\varrho, \varphi, p, q, r, s )^{\mathrm{T}} ~\in X,
\label{canonical_variables}
\end{equation}
where $\xi_1=\varrho$, $\xi_3=p$, $\xi_5={r}$ are $3$-forms and
$\xi_2=\varphi$, $\xi_4=q,$ $\xi_6 ={s}$ are 0-forms (scalars) on the base space ${M}=\mathbb{T}^3$.  
Thus half of the components of $X$ are 0-forms and half 3-forms. 
We denote by $X^*$ the linear space dual of $X$;
the odd number components of $\bm{\eta}\in X^*$ are 0-forms and the even number components are 3-forms.
The pairing of $X^*$ and $X$, formally given by (7), is 
\begin{equation}
\langle \bm{\eta}, \bm{\xi} \rangle = \sum_{j} \int_{M} \eta_j \wedge \xi_j ,
\quad \bm{\eta}\in X^*, ~ \bm{\xi}\in X.
\label{pairing}
\end{equation}
We will denote by $\alpha^*$ the Hodge dual of a differential form $\alpha$.

The relation between the physical quantity $\bm{u}\in \mathcal{V}$ and the Clebsch parameters $\bm{\xi}\in X$ are
\begin{equation}
\rho\Leftrightarrow \varrho^*,
\label{density}
\end{equation}
(i.e., $\varrho^* \vol^3 = \varrho$ with the volume $3$-form $\vol^3$),
and, introducing scalars $\check{p}={p^*}/{\varrho^*}$ and $\check{{r}}={{r}^*}/{\varrho^*}$,
\begin{equation}
\bm{V} \Leftrightarrow    \wp = \rmd\varphi + \check{p} \rmd q + \check{{r}} \rmd {s} 
\label{Clebsch}
\end{equation}
The Hamiltonian of (11) reads
\begin{equation}
H(\bm{\xi}) = \int_{M} \left[ \frac{1}{2} {\left| \rmd\varphi + \frac{p^*}{\varrho^*} \rmd q 
+\frac{{r}^*}{\varrho^*} \rmd{s}\right|^2} + \varepsilon(\varrho^*)  \right] \varrho ,
\label{Hamiltonian}
\end{equation}
where $|v|^2=v\wedge*v/\vol^n$ for $p$-form $v$.
The helicity now reads $C=\int \wp\wedge\rmd\wp$;
notice that the helicity density $\wp\wedge\rmd\wp$ is a 3-form
(which does not have a proper definition in a 2D space).

The canonical Hamiltonian system (12) has a profound geometrical meaning.
Let $\mathcal{L}_{\bm{V}}$ be the conventional Lie derivative along the vector $\bm{V}\in T{M}$, i.e.,
by Cartan's formula,
\begin{equation}
\mathcal{L}_{\bm{V}} = \rmd i_{\bm{V}} +  i_{\bm{V}} \rmd.
\label{Cartan}
\end{equation}
See Table\,\ref{table:translation}, for the translation of differential-geometry notation and vector-analysis notation.
It is convenient to add the time axis to $M$ and define the Lie derivative for the 4-velocity $\tilde{\bm{V}}=(1,\bm{V})^{\mathrm{T}}$
(which is readily generalized to a relativistic 4-velocity when we formulate a relativistic model);
we denote $\widetilde{\mathcal{L}}_{\tilde{\bm{V}}} = \partial_t + \mathcal{L}_{\bm{V}}$.
Now the system (12) reads
\begin{equation}
\left\{
\begin{array}{l}
\widetilde{\mathcal{L}}_{\tilde{\bm{V}}} \varrho = 0, \\
\widetilde{\mathcal{L}}_{\tilde{\bm{V}}} \varphi = h- \frac{1}{2}{V^2},
\end{array} \right.
\quad
\left\{
\begin{array}{l}
\widetilde{\mathcal{L}}_{\tilde{\bm{V}}} p = 0, \\
\widetilde{\mathcal{L}}_{\tilde{\bm{V}}} q = 0,
\end{array} \right.
\quad
\left\{
\begin{array}{l}
\widetilde{\mathcal{L}}_{\tilde{\bm{V}}} {r} = 0 , \\
\widetilde{\mathcal{L}}_{\tilde{\bm{V}}} {s} = 0 .
\end{array} \right.
\label{H-1} 
\end{equation}
Notice that all Clebsch variables, excepting the scalar $\varphi$, are just `Lie-dragged' by the flow velocity $\bm{V}$.
The `nonlinearity' of fluid mechanics comes from the right-hand-side term $ h- V^2/2$ on the
transport equation of $\varphi$.

Epi-2D flow is a reduction of the general 3D flow by setting ${r}=0$.  
The corresponding physical fields are $\rho\Leftrightarrow \varrho^*$ and 
\begin{equation}
\bm{V} \Leftrightarrow  \wp =  \rmd\varphi + \check{p} \rmd q ,
\quad \left(\check{p}= {p^*}/{\varrho^*}\right) .
\label{epi-2D-2}
\end{equation}
The reduced Hamiltonian reads
\begin{equation}
H(\bm{\xi}) = \int_{M} \left[ \frac12 {\left| \rmd\varphi + \frac{p^*}{\varrho^*} \rmd q \right|^2 }  + \varepsilon(\varrho^*)  \right] \varrho\, .
\label{reduces-Hamiltonian}
\end{equation}

Epi-2D flow may have a finite vorticity $\bm{\omega}=\rmd\wp=\rmd\check{p}\wedge\rmd q$, but its 
helicity density
$\wp\wedge\rmd\wp = \rmd\varphi\wedge \rmd\check{p}\wedge\rmd q = \rmd(\varphi\wedge \rmd\check{p}\wedge\rmd q )$ is an exact 3-form, implying zero helicity,  $C=\int_{{M}} \wp\wedge\rmd\wp =0$.

With the phantom (scalar) $s$, the generalized enstrophy (16) reads
\begin{equation}
{Q} := \int_{\Omega(t)} f(\vartheta)\, \varrho,
\quad \vartheta = \frac{(\omega \wedge \rmd{s})^*}{\varrho^*} .
\label{enstrophy-2}
\end{equation}
The key to find such a constant of motion is to combine the Lie-dragged forms $\varrho,\, p,\, q,\, r$, and $s$ to constitute a 3-form.

\vspace*{1cm}
\section{Morphology}

Here we describe some examples for visualizing  epi-2D flows, along with the  fusion of epi-2D flows that  generate 3D flows.

Separating out  the component of potential flow, we consider epi-2D flows represented by 
\begin{equation}
\bm{V}_1 = p \nabla q, \quad \bm{V}_2 = r\nabla s,
\label{rep}
\end{equation}
where we invoke  conventional vector-analysis notation.
The  vorticities corresponding to \eqref{rep} are
\[
\bm{\omega}_1 =\nabla p \times \nabla q, \quad \bm{\omega}_2 =\nabla r \times \nabla s.
\]
With a co-moving scalar $\tau$, we define the charge densities (generalized enstropies) 
\[
\mathscr{Q}_+ = \nabla \tau \cdot \bm{\omega}_1, \quad
\mathscr{Q}_- = \nabla \tau \cdot \bm{\omega}_2 .
\]
Upon choosing the  reference scalar $\tau$ to be  $\tau=s$ for $\mathscr{Q}_+$ and  $\tau=q$ for $\mathscr{Q}_-$,  the residual helicity density reads
\[
\mathscr{C} = r \mathscr{Q}_+ + p \mathscr{Q}_-.
\]

Concentrating  on localized flows that look like \emph{particles},  we assume $p$ and $r$ have  Gaussian-like factors ($\exp(-t^2)$)  or mollifier  factors ($\exp[-1/(1-|t|^2)]$ for $|t|<1$, continued to $0$ for $|t|\geq1$).


\begin{figure*}[h]
\label{fig:epi-2D-0}
\vspace*{3cm}
\begin{center}
\raisebox{4cm}{\textbf{(a)}}
\includegraphics[scale=0.4]{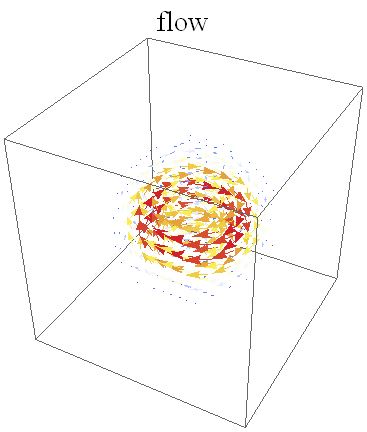} ~~
\includegraphics[scale=0.4]{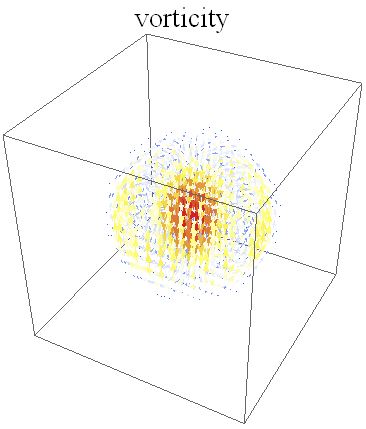}~~
\includegraphics[scale=0.4]{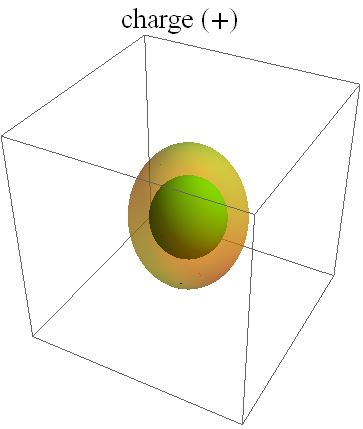}
\\ 
$\underline{~~~~~~~~~~~~~~~~~~~~~~~~~~~~~~~~~~~~~~~~~~~~~~~~~~~~~~~~~~~~~~~~~~~~~~~~~~~~~~~~~~~~~~~~~~~~}$
\\  ~ \\
\raisebox{4cm}{\textbf{(b)}}
\includegraphics[scale=0.4]{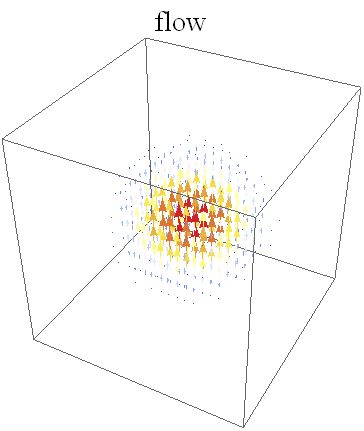} ~~
\includegraphics[scale=0.4]{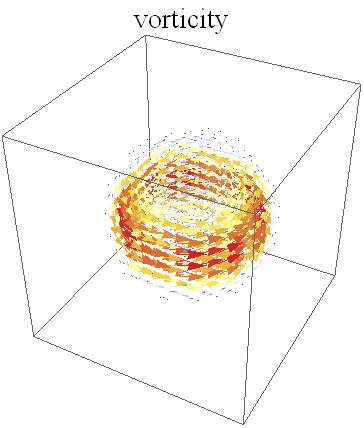} ~~
\includegraphics[scale=0.4]{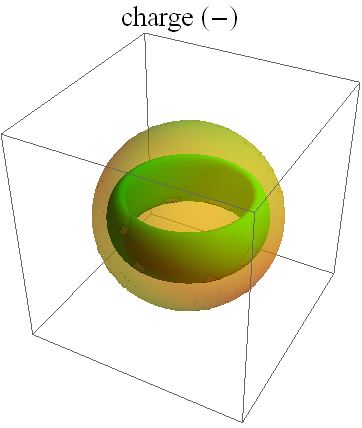}
\\ 
$\underline{~~~~~~~~~~~~~~~~~~~~~~~~~~~~~~~~~~~~~~~~~~~~~~~~~~~~~~~~~~~~~~~~~~~~~~~~~~~~~~~~~~~~~~~~~~~~}$
\\  ~ \\
\raisebox{4cm}{\textbf{(c)}}
\includegraphics[scale=0.4]{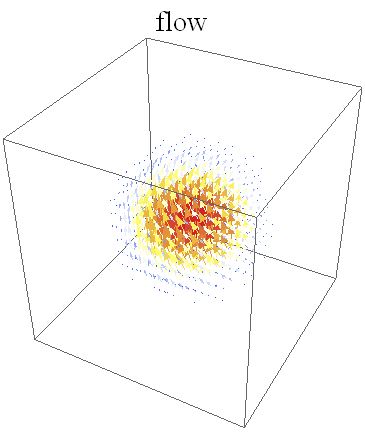} ~~
\includegraphics[scale=0.4]{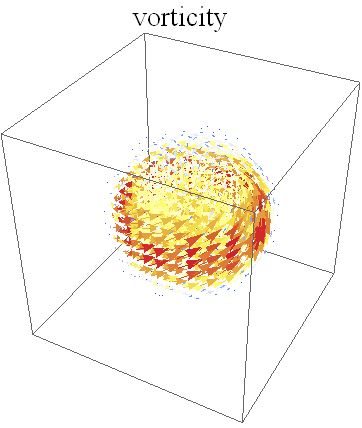} ~~
\includegraphics[scale=0.4]{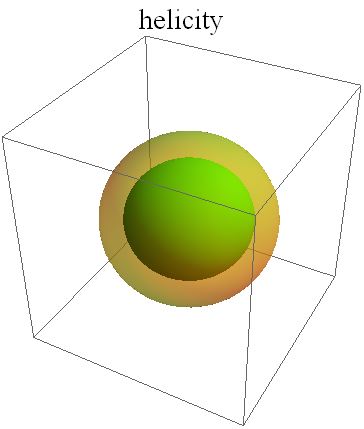}
\\ ~ \\
\caption{
(a) An example of localized epi-2D flow with  $\bm{V}_1=p \nabla q$, where 
$p= (x^2 + y^2) \phi(\sqrt{x^2 + y^2 + z^2})$  and $q = \arctan (x/y)$,
which gives a primarily circulating flow. 
Here $\phi(\tau) = \exp [-1/(1-|\tau|^2) ]$ for $|\tau|< 1$, while  $\phi(\tau)=0$ for $|\tau|\geq 1$,  using   the Friedrichs mollifier. 
The contours show the levels  $\mathscr{Q}_+= -0.1$ and $-0.5$.
(b) An example of localized epi-2D flow with  $\bm{V}_2= r \nabla s$, where  
$r= \phi(\sqrt{x^2 + y^2 + z^2})$ and $s= \tanh z$, which gives a primarily longitudinal flow.
The contours show the levels  $\mathscr{Q}_-= -0.1$ and $-1$.
(c) An example of localized 3D flow that is generated by the fusion (superposition) of $\bm{V}_1$ and $\bm{V}_2$, generating a  finite helicity density.
The contours show the levels  $\mathscr{C}= -0.01$ and $-0.1$.
}
\end{center}
\end{figure*}

\begin{figure*}[h]
\label{fig:epi-2D-deform}
\vspace*{3cm}
\begin{center}
\raisebox{4cm}{\textbf{(a)}}
\includegraphics[scale=0.4]{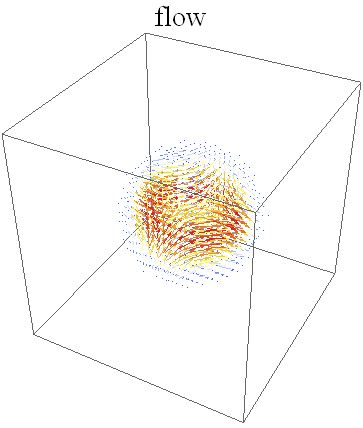} ~~
\includegraphics[scale=0.4]{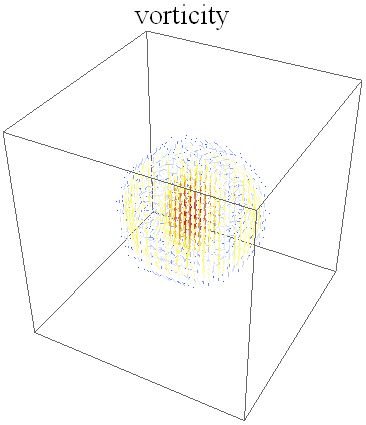}~~
\includegraphics[scale=0.4]{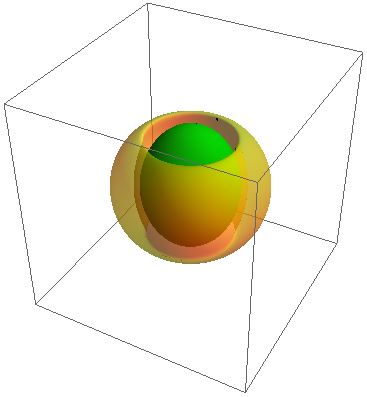}
\\ 
$\underline{~~~~~~~~~~~~~~~~~~~~~~~~~~~~~~~~~~~~~~~~~~~~~~~~~~~~~~~~~~~~~~~~~~~~~~~~~~~~~~~~~~~~~~~~~~~~}$
\\  ~ \\
\raisebox{4cm}{\textbf{(b)}}
\includegraphics[scale=0.4]{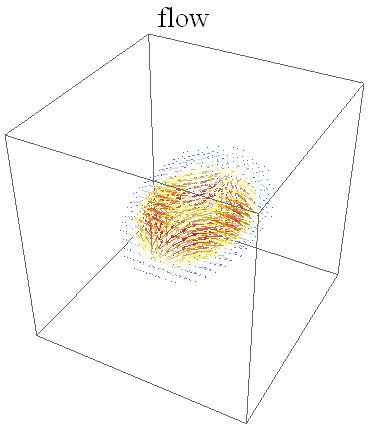} ~~
\includegraphics[scale=0.4]{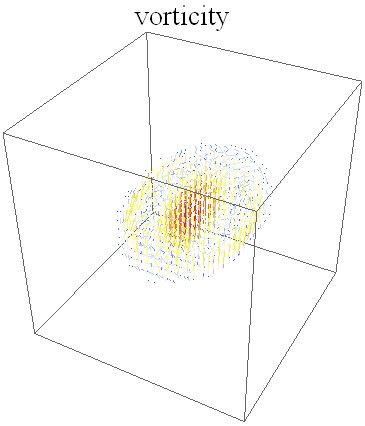}~~
\includegraphics[scale=0.4]{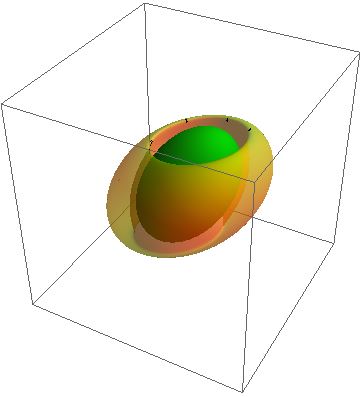}
\\ 
$\underline{~~~~~~~~~~~~~~~~~~~~~~~~~~~~~~~~~~~~~~~~~~~~~~~~~~~~~~~~~~~~~~~~~~~~~~~~~~~~~~~~~~~~~~~~~~~~}$
\\  ~ \\
\raisebox{4cm}{\textbf{(c)}}
\includegraphics[scale=0.4]{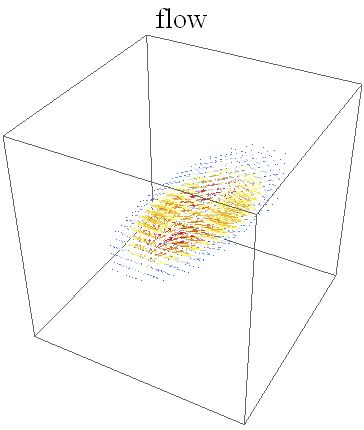} ~~
\includegraphics[scale=0.4]{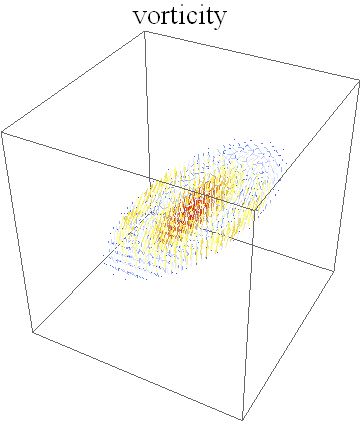}~~
\includegraphics[scale=0.4]{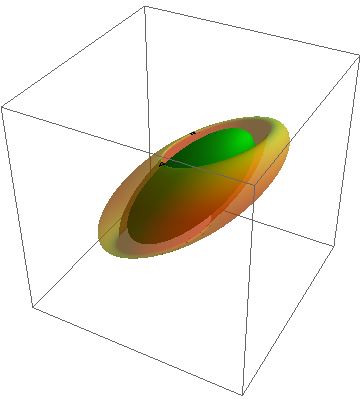}
\\ ~ \\
\caption{
Deformation by a potential flow $\bm{V}_0=\nabla (yz)$ of an epi-2D particle
$\bm{V}_1=p \nabla q$ with
(a) the initial condition
$p= (x^2 + y^2) \phi(\sqrt{x^2 + y^2 + z^2})$ and $q = \arctan (x/y)$.
The contours  $\mathscr{Q}_+= \pm 0.1$ are shown.
(b) The deformed particle at $t=0.2$.
Here we assume that $p$ and $q$ are passive scalars transported by $\bm{V}_0$.
The deformation is written as 
$y \mapsto (\cosh t)y  -(\sinh t) z$, and
$z \mapsto -(\sinh t)y +( \cosh t) z$.
(c) The deformed particle at $t=0.5$.
}
\end{center}
\end{figure*}

\begin{figure*}[tb]
\label{fig:epi-2D-fusion-cv}
\vspace*{3cm}
\begin{center}
$~~~~~~\bm{V} ~~~~~~~~~~~~~~~~~~~~~~\bm{\omega} ~~~~~~~~~~~~~~~~~~~~\mathscr{Q}_+  ~~~~~~~~~~~~~~~~~~~\mathscr{Q}_-  ~~~~~~~~~~~~~~~~~~\mathscr{C}$
\\
$\underline{~~~~~~~~~~~~~~~~~~~~~~~~~~~~~~~~~~~~~~~~~~~~~~~~~~~~~~~~~~~~~~~~~~~~~~~~~~~~~~~~~~~~~~~~~~~~~~~~~~~~~~~~~~~}$
\\ ~ \\
\raisebox{3cm}{\textbf{(a)}}~~
\includegraphics[scale=0.28]{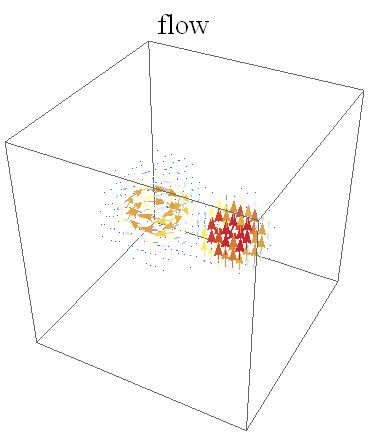} ~
\includegraphics[scale=0.28]{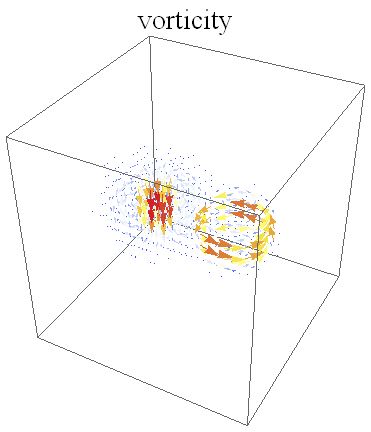}~
\includegraphics[scale=0.28]{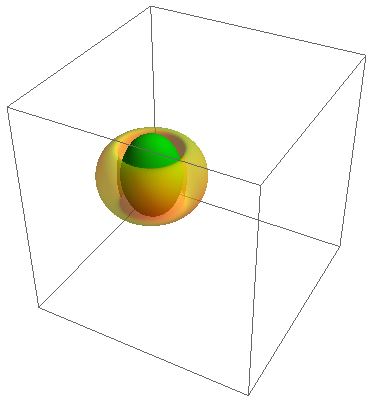}~
\includegraphics[scale=0.28]{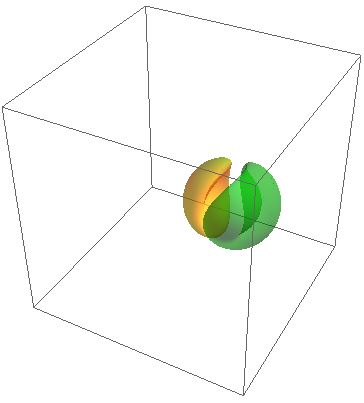}~
\includegraphics[scale=0.28]{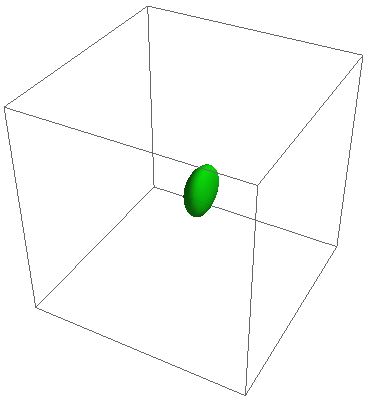}
\\ 
$\underline{~~~~~~~~~~~~~~~~~~~~~~~~~~~~~~~~~~~~~~~~~~~~~~~~~~~~~~~~~~~~~~~~~~~~~~~~~~~~~~~~~~~~~~~~~~~~~~~~~~~~~~~~~~~}$
\\  ~ \\
\raisebox{3cm}{\textbf{(b)}}~~
\includegraphics[scale=0.28]{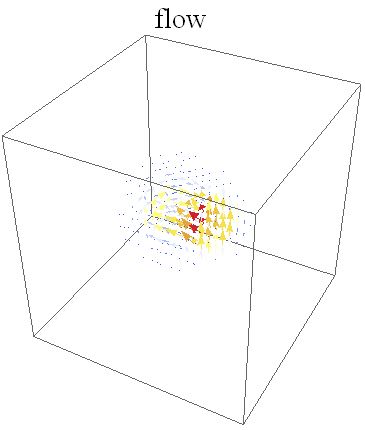}~
\includegraphics[scale=0.28]{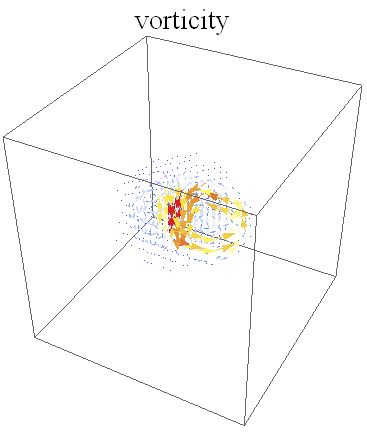}~
\includegraphics[scale=0.28]{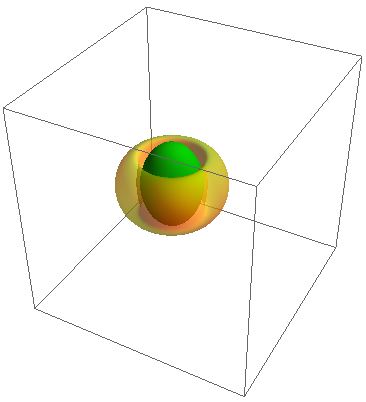}~
\includegraphics[scale=0.28]{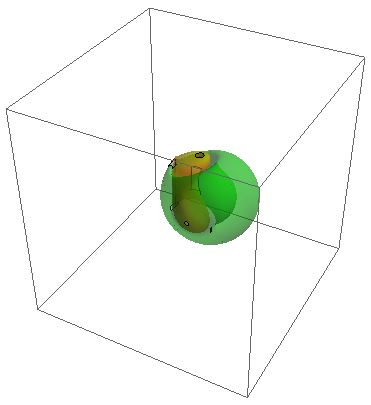}~
\includegraphics[scale=0.28]{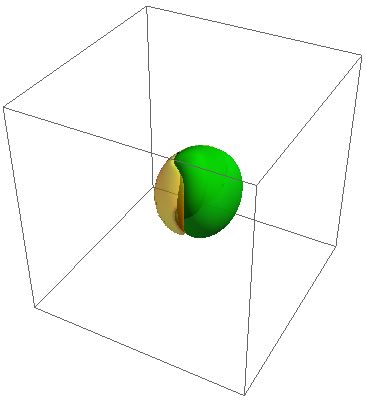}
\\ 
$\underline{~~~~~~~~~~~~~~~~~~~~~~~~~~~~~~~~~~~~~~~~~~~~~~~~~~~~~~~~~~~~~~~~~~~~~~~~~~~~~~~~~~~~~~~~~~~~~~~~~~~~~~~~~~~}$
\\  ~ \\
\raisebox{3cm}{\textbf{(c)}}~~
\includegraphics[scale=0.28]{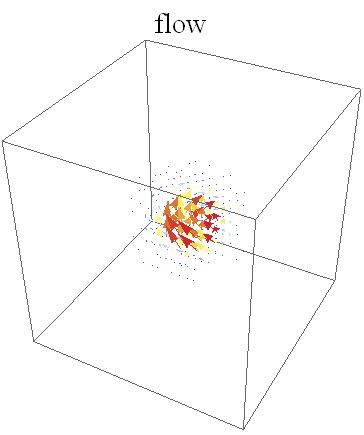}~
\includegraphics[scale=0.28]{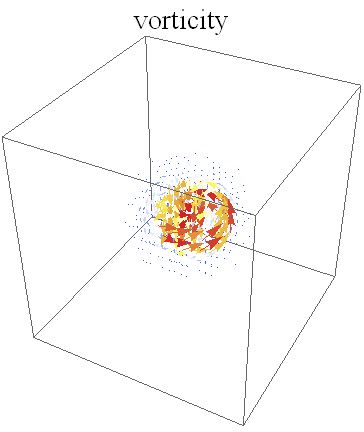}~
\includegraphics[scale=0.28]{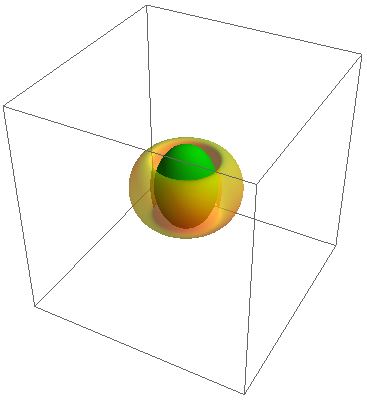}~
\includegraphics[scale=0.28]{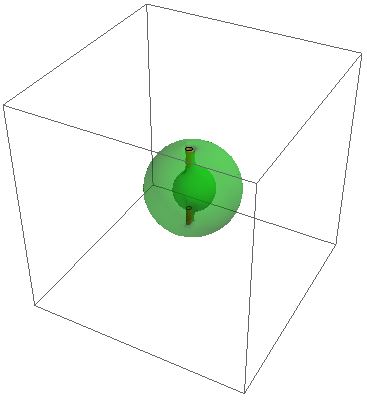}~
\includegraphics[scale=0.28]{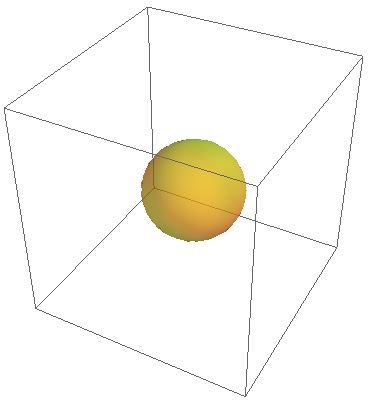}
\\
$\underline{~~~~~~~~~~~~~~~~~~~~~~~~~~~~~~~~~~~~~~~~~~~~~~~~~~~~~~~~~~~~~~~~~~~~~~~~~~~~~~~~~~~~~~~~~~~~~~~~~~~~~~~~~~~}$
\\  ~ \\
\raisebox{3cm}{\textbf{(d)}}~~
\includegraphics[scale=0.28]{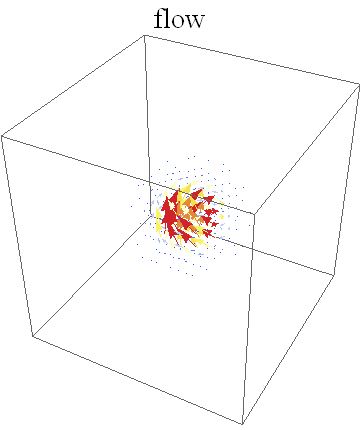}~
\includegraphics[scale=0.28]{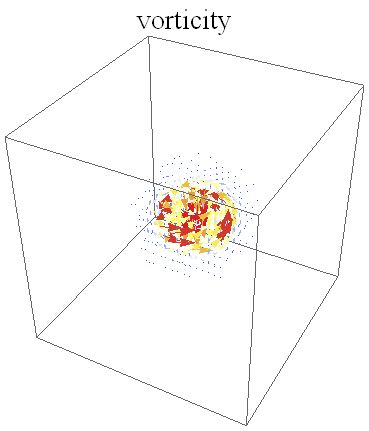}~
\includegraphics[scale=0.28]{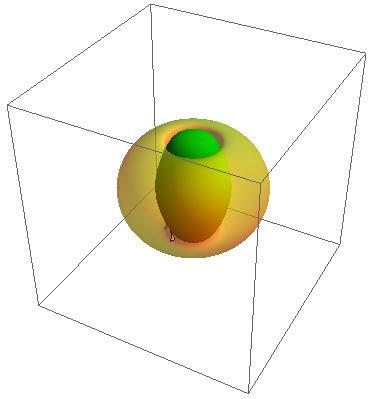}~
\includegraphics[scale=0.28]{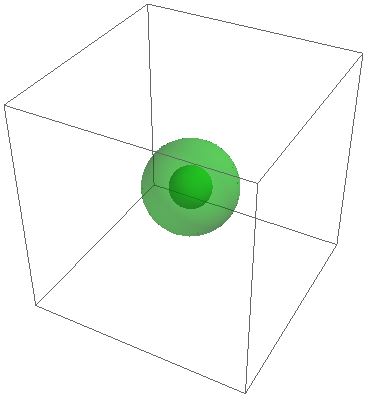}~
\includegraphics[scale=0.28]{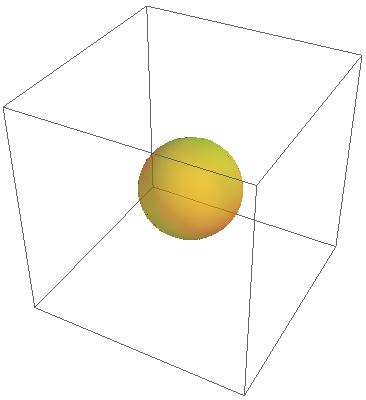}
\\ 
\caption{
Fusion of circulating and longitudinal epi-2D particles with 
$p= \alpha (x_1^2 + y^2) \rme^{-(x_1^2 + y^2 + z^2)^2}$, 
$q = \arctan (x_1/y)$,
$r=\rme^{-(x_2^2 + y^2 + z^2)^4 } $, and 
$s= \tanh (z/2) $,
where $x_1=x+\delta$, $x_2=x-\delta$.
Here   $\alpha=0.8$ was chosen to balance the magnitudes of both charges.
The merging parameter $\delta$ is chosen in (a) $\delta=1.2$, (b) $\delta=0.5$, (c) $\delta=0.1$, (d) $\delta=0$. 
Contours  $\mathscr{Q}_\pm=\pm0.1$ and $\mathscr{C}_{\mathrm{r}}=10^{-3}$ (for $\delta=1.2$), $\pm10^{-3}$ (for $\delta=0.5$),
$-10^{-3}$ (for $\delta=0.1$ and $0$) are shown.
}
\end{center}
\end{figure*}

\begin{figure*}[tb]
\label{fig:epi-2D-fusion-cc}
\vspace*{3cm}
\begin{center}
$~~~~~~\bm{V} ~~~~~~~~~~~~~~~~~~~~~~\bm{\omega} ~~~~~~~~~~~~~~~~~~~~\mathscr{Q}_+  ~~~~~~~~~~~~~~~~~~~\mathscr{Q}_-  ~~~~~~~~~~~~~~~~~~\mathscr{C}$
\\ ~ \\
$\underline{~~~~~~~~~~~~~~~~~~~~~~~~~~~~~~~~~~~~~~~~~~~~~~~~~~~~~~~~~~~~~~~~~~~~~~~~~~~~~~~~~~~~~~~~~~~~~~~~~~~~~~~~~~~}$
\\
\raisebox{3cm}{\textbf{(a)}}~~
\includegraphics[scale=0.28]{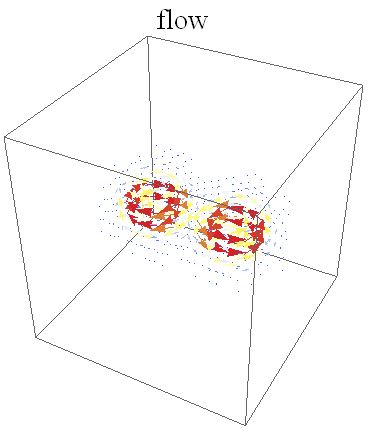} ~
\includegraphics[scale=0.28]{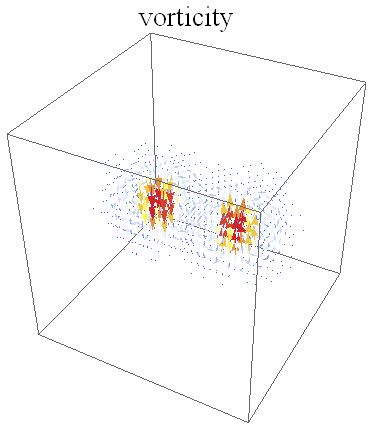}~
\includegraphics[scale=0.28]{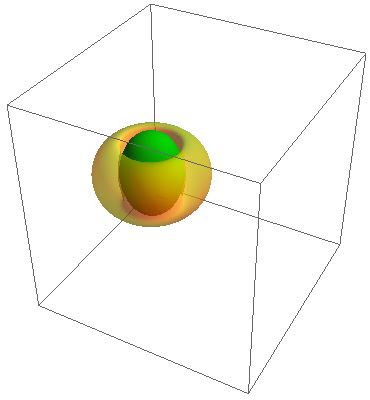}~
\includegraphics[scale=0.28]{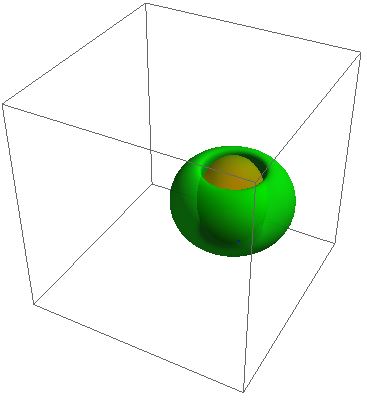}~
\includegraphics[scale=0.28]{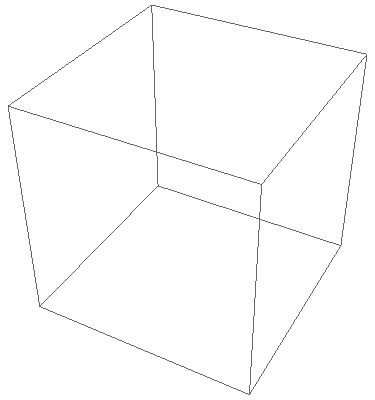}
\\ 
$\underline{~~~~~~~~~~~~~~~~~~~~~~~~~~~~~~~~~~~~~~~~~~~~~~~~~~~~~~~~~~~~~~~~~~~~~~~~~~~~~~~~~~~~~~~~~~~~~~~~~~~~~~~~~~~}$
\\  ~ \\
\raisebox{3cm}{\textbf{(b)}}~~
\includegraphics[scale=0.28]{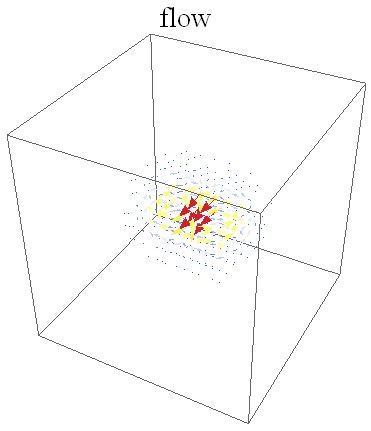}~
\includegraphics[scale=0.28]{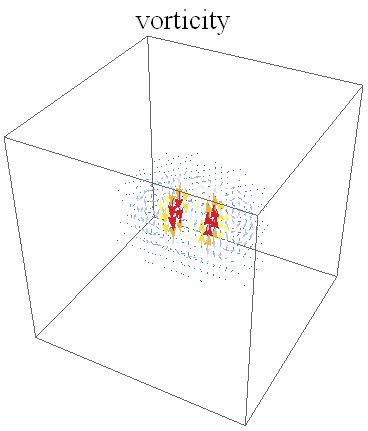}~
\includegraphics[scale=0.28]{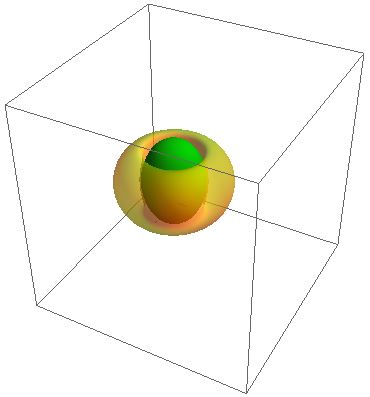}~
\includegraphics[scale=0.28]{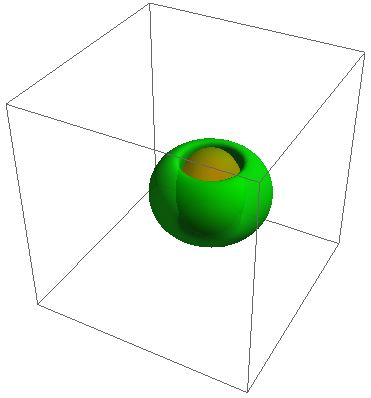}~
\includegraphics[scale=0.28]{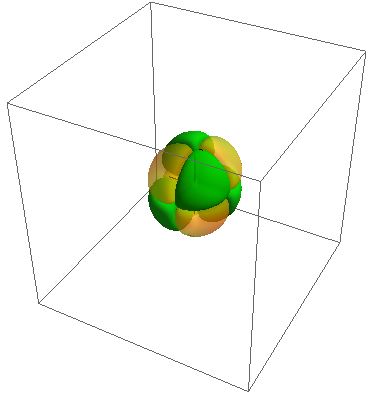}
\\ 
$\underline{~~~~~~~~~~~~~~~~~~~~~~~~~~~~~~~~~~~~~~~~~~~~~~~~~~~~~~~~~~~~~~~~~~~~~~~~~~~~~~~~~~~~~~~~~~~~~~~~~~~~~~~~~~~}$
\\  ~ \\
\raisebox{3cm}{\textbf{(c)}}~~
\includegraphics[scale=0.28]{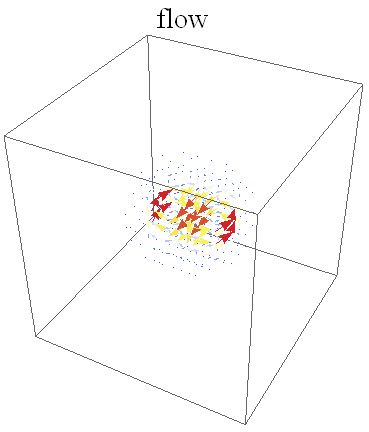}~
\includegraphics[scale=0.28]{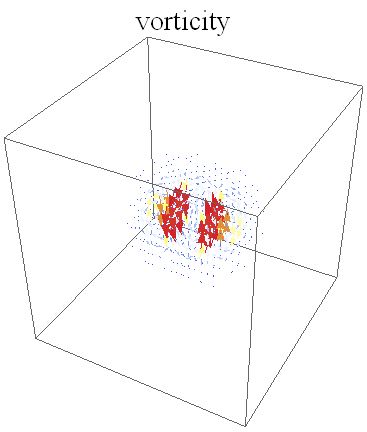}~
\includegraphics[scale=0.28]{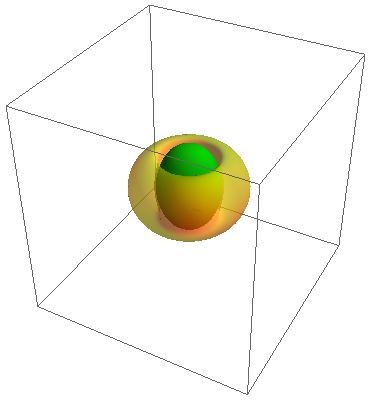}~
\includegraphics[scale=0.28]{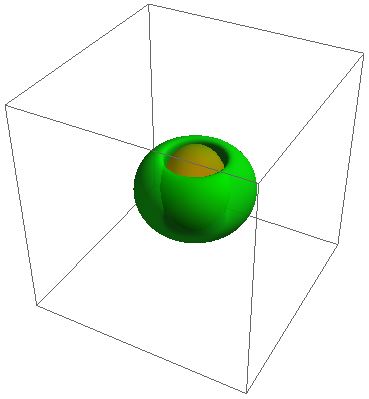}~
\includegraphics[scale=0.28]{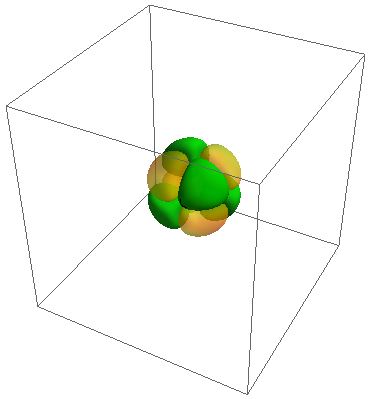}
\\
$\underline{~~~~~~~~~~~~~~~~~~~~~~~~~~~~~~~~~~~~~~~~~~~~~~~~~~~~~~~~~~~~~~~~~~~~~~~~~~~~~~~~~~~~~~~~~~~~~~~~~~~~~~~~~~~}$
\\  ~ \\
\raisebox{3cm}{\textbf{(d)}}~~
\includegraphics[scale=0.28]{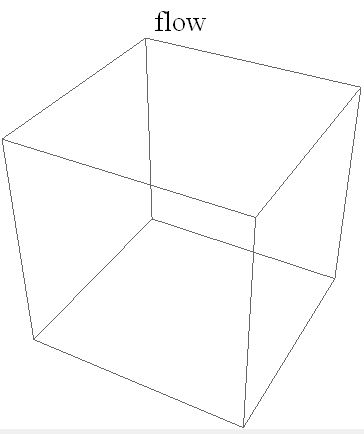}~
\includegraphics[scale=0.28]{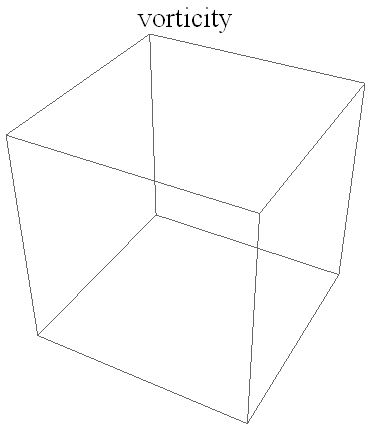}~
\includegraphics[scale=0.28]{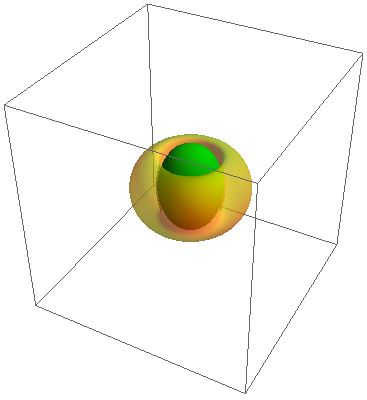}~
\includegraphics[scale=0.28]{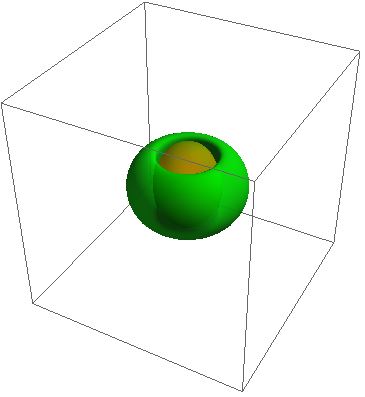}~
\includegraphics[scale=0.28]{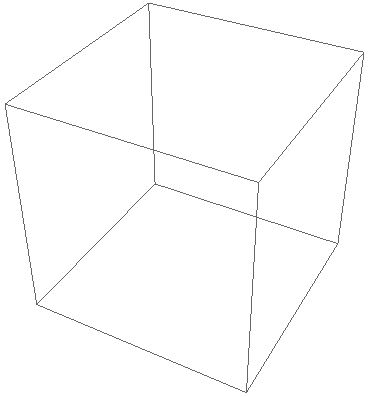}
\\ 
\caption{
The fusion of a pair of epi-2D particles that circulate in the opposite directions with 
$p=  (x_1^2 + y^2) \rme^{-(x_1^2 + y^2 + z^2)^4}$, 
$q = \arctan (x_1/y)$,
$r= \alpha (x_2^2 + y^2) \rme^{-(x_2^2 + y^2 + z^2)^4 } $, and 
$s= -\arctan (x_2/y) $,
where $x_1=x+\delta$, $x_2=x-\delta$.
The charges are evaluated as
$\mathscr{Q}_+ = \rmd p \wedge\rmd q\wedge\rmd \tau$ and
$\mathscr{Q}_- = \rmd r \wedge\rmd s\wedge\rmd \tau$
with $\tau = \tanh z$.
The merging parameter $\delta$ is chosen in (a) $\delta=1.2$, (b) $\delta=0.5$, (c) $\delta=0.1$, (d) $\delta=0$. 
Contours  $\mathscr{Q}_\pm=\pm0.1$ and $\mathscr{C}_{\mathrm{r}}=\pm 10^{-3}$ are shown.
}
\end{center}
\end{figure*}

\begin{figure*}[tb]
\label{fig:epi-2D-fusion-c-c}
\vspace*{3cm}
\begin{center}
$~~~~~~\bm{V} ~~~~~~~~~~~~~~~~~~~~~~\bm{\omega} ~~~~~~~~~~~~~~~~~~~~\mathscr{Q}_+  ~~~~~~~~~~~~~~~~~~~\mathscr{Q}_-  ~~~~~~~~~~~~~~~~~~\mathscr{C}$
\\ ~ \\
$\underline{~~~~~~~~~~~~~~~~~~~~~~~~~~~~~~~~~~~~~~~~~~~~~~~~~~~~~~~~~~~~~~~~~~~~~~~~~~~~~~~~~~~~~~~~~~~~~~~~~~~~~~~~~~~}$
\\
\raisebox{3cm}{\textbf{(a)}}~~
\includegraphics[scale=0.28]{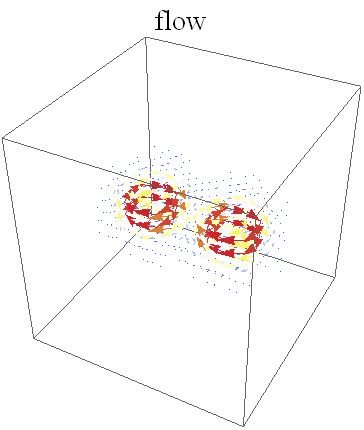} ~
\includegraphics[scale=0.28]{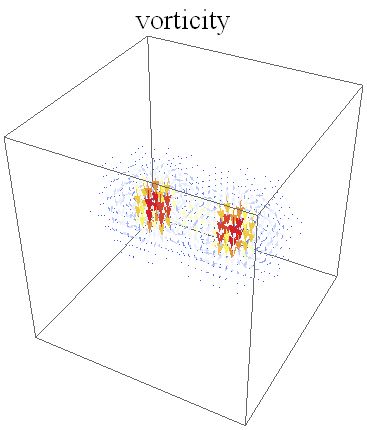}~
\includegraphics[scale=0.28]{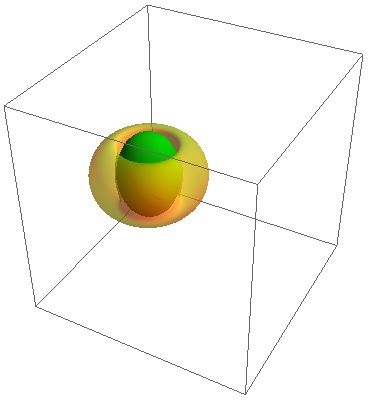}~
\includegraphics[scale=0.28]{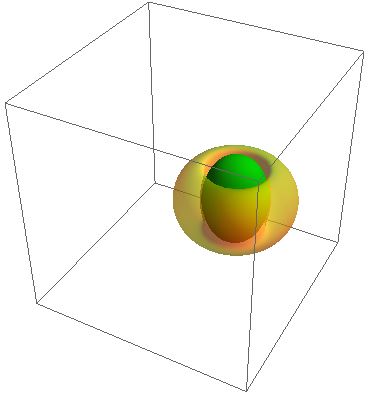}~
\includegraphics[scale=0.28]{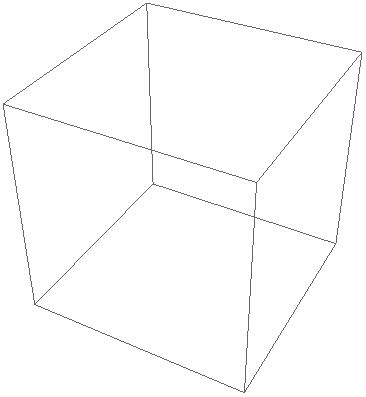}
\\ 
$\underline{~~~~~~~~~~~~~~~~~~~~~~~~~~~~~~~~~~~~~~~~~~~~~~~~~~~~~~~~~~~~~~~~~~~~~~~~~~~~~~~~~~~~~~~~~~~~~~~~~~~~~~~~~~~}$
\\  ~ \\
\raisebox{3cm}{\textbf{(b)}}~~
\includegraphics[scale=0.28]{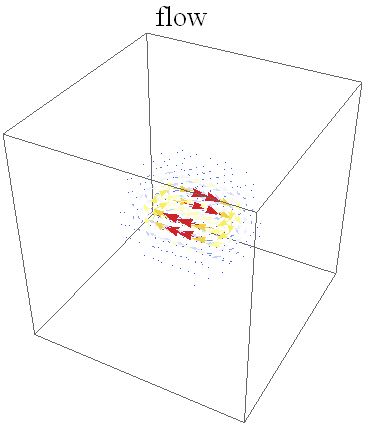}~
\includegraphics[scale=0.28]{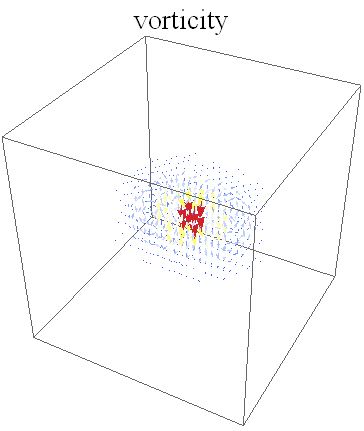}~
\includegraphics[scale=0.28]{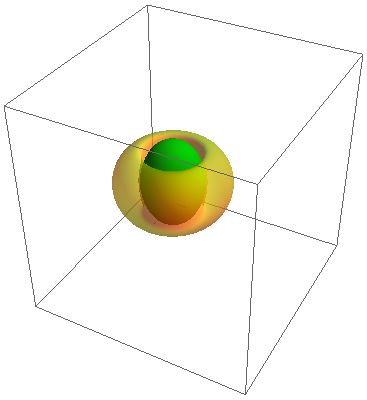}~
\includegraphics[scale=0.28]{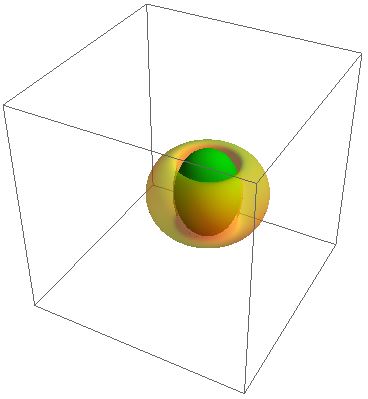}~
\includegraphics[scale=0.28]{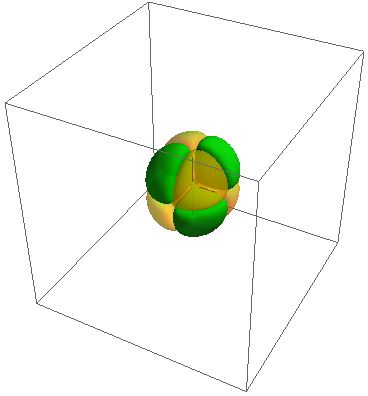}
\\ 
$\underline{~~~~~~~~~~~~~~~~~~~~~~~~~~~~~~~~~~~~~~~~~~~~~~~~~~~~~~~~~~~~~~~~~~~~~~~~~~~~~~~~~~~~~~~~~~~~~~~~~~~~~~~~~~~}$
\\  ~ \\
\raisebox{3cm}{\textbf{(c)}}~~
\includegraphics[scale=0.28]{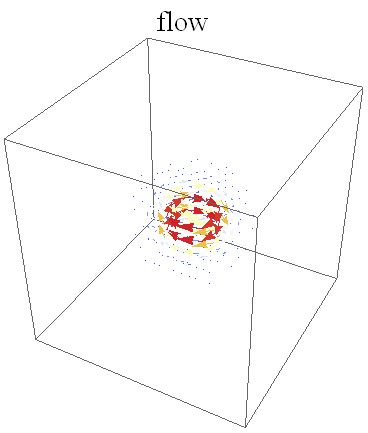}~
\includegraphics[scale=0.28]{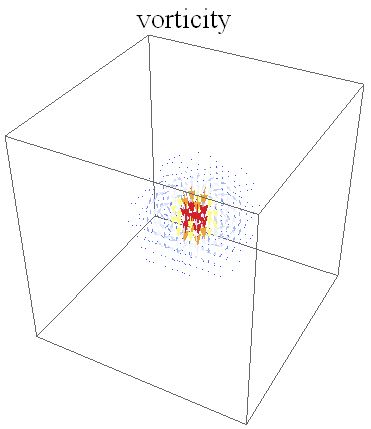}~
\includegraphics[scale=0.28]{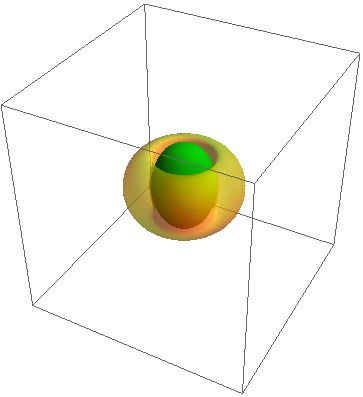}~
\includegraphics[scale=0.28]{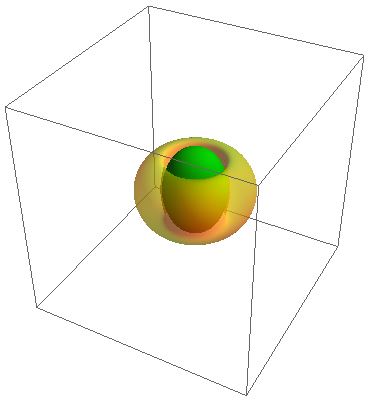}~
\includegraphics[scale=0.28]{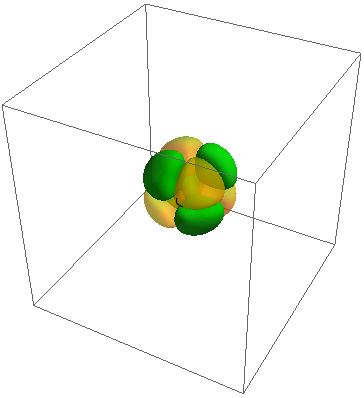}
\\
$\underline{~~~~~~~~~~~~~~~~~~~~~~~~~~~~~~~~~~~~~~~~~~~~~~~~~~~~~~~~~~~~~~~~~~~~~~~~~~~~~~~~~~~~~~~~~~~~~~~~~~~~~~~~~~~}$
\\  ~ \\
\raisebox{3cm}{\textbf{(d)}}~~
\includegraphics[scale=0.28]{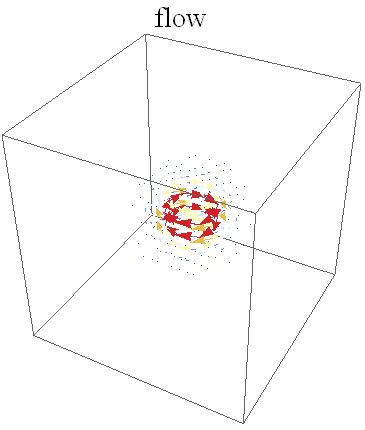}~
\includegraphics[scale=0.28]{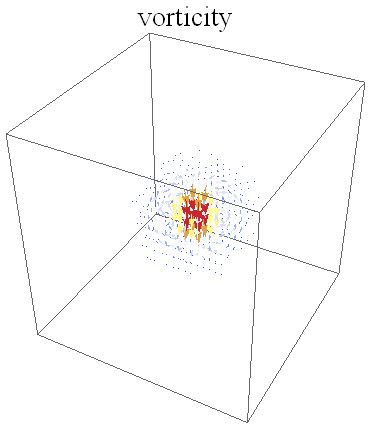}~
\includegraphics[scale=0.28]{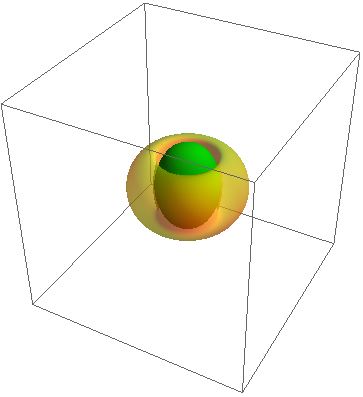}~
\includegraphics[scale=0.28]{fusion-c-c/fusion-c-c0_q1_q2}~
\includegraphics[scale=0.28]{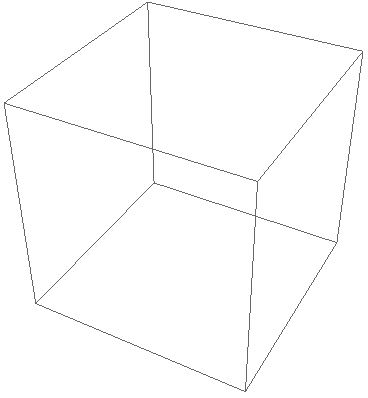}
\\ 
\caption{
Fusion of a pair of epi-2D particles that  circulate in the same directions with 
$p=  (x_1^2 + y^2) \rme^{-(x_1^2 + y^2 + z^2)^4}$, 
$q = \arctan (x_1/y)$,
$r= \alpha (x_2^2 + y^2) \rme^{-(x_2^2 + y^2 + z^2)^4 } $, and 
$s= \arctan (x_2/y) $,
where $x_1=x+\delta$, $x_2=x-\delta$.
The charges are evaluated as
$\mathscr{Q}_+ = \rmd p \wedge\rmd q\wedge\rmd \tau$ and
$\mathscr{Q}_- = \rmd r \wedge\rmd s\wedge\rmd \tau$
with $\tau = \tanh z$.
The merging parameter $\delta$ is chosen in (a) $\delta=1.2$, (b) $\delta=0.5$, (c) $\delta=0.1$, (d) $\delta=0$. 
Contours of $\mathscr{Q}_\pm=\pm0.1$ and $\mathscr{C}_{\mathrm{r}}=\pm 10^{-3}$ are shown.
As $p=\alpha r$ ($\alpha=$ constant) at $\delta=0$, 
the fusion of $\mathscr{Q}_+$ and $\mathscr{Q}_-$ yields zero helicity.
}
\end{center}
\end{figure*}


\begin{thebibliography}{99.}
\bibitem{suppl1}
A deeper mathematical structure is elucidated by formulating the system in terms of differential forms;
see Appendix B.
\bibitem{symmetries}
An axisymmetric flow such that $\partial_\theta=0$ and $\bm{e}_\theta\cdot\bm{V}=0$ is also 2D, in which
we may choose $s=\theta$, and then, the enstrophy of 
$\omega_\theta=r\nabla\theta\cdot\bm{\omega}$
is constant.
A helically symmetric flow
such that $\iota\partial_\theta -\partial_z=0$ and $\bm{h}\cdot\bm{V}=0$
($\bm{h}=\iota r\bm{e}_\theta - \bm{e}_z$ with a constant $\iota$)
is epi-2D,
which can be written as $\rmd \varphi + \check{p}\rmd\psi$ 
with $\varphi$ and $\check{p}$ being functions of $r$ and $\psi=m(\theta+\iota z)$.
It is not 2D, because $\bm{h}$ is not integrable to define a perpendicular 2D manifold,
however, it is very close to 2D in that the vortex stretching vanishes; 
cf.\,\cite{2.5D}.
\end{thebibliography}
\end{document}